\documentclass[article]{jss}\usepackage[]{graphicx}\usepackage[]{xcolor}
\makeatletter
\def\maxwidth{ %
  \ifdim\Gin@nat@width>\linewidth
    \linewidth
  \else
    \Gin@nat@width
  \fi
}
\makeatother

\definecolor{fgcolor}{rgb}{0.345, 0.345, 0.345}
\newcommand{\hlnum}[1]{\textcolor[rgb]{0.686,0.059,0.569}{#1}}%
\newcommand{\hlsng}[1]{\textcolor[rgb]{0.192,0.494,0.8}{#1}}%
\newcommand{\hlcom}[1]{\textcolor[rgb]{0.678,0.584,0.686}{\textit{#1}}}%
\newcommand{\hlopt}[1]{\textcolor[rgb]{0,0,0}{#1}}%
\newcommand{\hldef}[1]{\textcolor[rgb]{0.345,0.345,0.345}{#1}}%
\newcommand{\hlkwa}[1]{\textcolor[rgb]{0.161,0.373,0.58}{\textbf{#1}}}%
\newcommand{\hlkwb}[1]{\textcolor[rgb]{0.69,0.353,0.396}{#1}}%
\newcommand{\hlkwc}[1]{\textcolor[rgb]{0.333,0.667,0.333}{#1}}%
\newcommand{\hlkwd}[1]{\textcolor[rgb]{0.737,0.353,0.396}{\textbf{#1}}}%

\usepackage{framed}
\makeatletter
\newenvironment{kframe}{%
 \def\at@end@of@kframe{}%
 \ifinner\ifhmode%
  \def\at@end@of@kframe{\end{minipage}}%
  \begin{minipage}{\columnwidth}%
 \fi\fi%
 \def\FrameCommand##1{\hskip\@totalleftmargin \hskip-\fboxsep
 \colorbox{shadecolor}{##1}\hskip-\fboxsep
     \hskip-\linewidth \hskip-\@totalleftmargin \hskip\columnwidth}%
 \MakeFramed {\advance\hsize-\width
   \@totalleftmargin\z@ \linewidth\hsize
   \@setminipage}}%
 {\par\unskip\endMakeFramed%
 \at@end@of@kframe}
\makeatother

\definecolor{shadecolor}{rgb}{.97, .97, .97}
\definecolor{messagecolor}{rgb}{0, 0, 0}
\definecolor{warningcolor}{rgb}{1, 0, 1}
\definecolor{errorcolor}{rgb}{1, 0, 0}
\newenvironment{knitrout}{}{} 

\usepackage{alltt}

\DeclarePairedDelimiter\parentheses{\lparen}{\rparen}
\DeclarePairedDelimiter\brackets{\lbrack}{\rbrack}

\newcommand{\bz}{\boldsymbol{z}}
\newcommand{\bZ}{\boldsymbol{Z}}
\newcommand{\bb}{\boldsymbol{\beta}}
\NewDocumentCommand{\bbh}{O{} O{}}{\widehat{\bb}_{#1}^{#2}}

\DeclareMathOperator{\nE}{\mathbb{E}}

\NewDocumentCommand{\mE}{O{} O{} O{}}{\nE_{#2}^{#3} \brackets*{#1}}

\NewDocumentCommand{\M}{O{} O{} O{}}{M_{#2}^{#3} \parentheses*{#1}}
\NewDocumentCommand{\Mh}{O{} O{} O{}}{\widehat{M}_{#2}^{#3} \parentheses*{#1}}

\newcommand{\W}{\boldsymbol{W}}
\newcommand{\Wh}{\widehat{\W}}
\NewDocumentCommand{\bW}{O{} O{} O{}}{\W_{#2}^{#3} \parentheses*{#1}}
\NewDocumentCommand{\bWh}{O{} O{} O{}}{\Wh_{#2}^{#3} \parentheses*{#1}}

\newcommand{\D}{\boldsymbol{D}}
\newcommand{\Dh}{\widehat{\D}}
\NewDocumentCommand{\bD}{O{} O{} O{}}{\D_{#2}^{#3} \parentheses*{#1}}
\NewDocumentCommand{\bDh}{O{} O{} O{}}{\Dh_{#2}^{#3} \parentheses*{#1}}

\NewDocumentCommand{\score}{O{} O{} O{}}{\boldsymbol{U}_{#2}^{#3} \parentheses*{#1}}

\DeclareMathOperator{\napp}{app}
\DeclareMathOperator{\nobs}{obs}

\newcommand{\mvar}[1]{\text{Var} \brackets*{#1}}

\usepackage{multirow}
\usepackage{array}

\usepackage[normalem]{ulem}
\usepackage{listings}
\author{\begin{tabular}{*{2}{>{\centering}p{.5\textwidth}}}
  \large Nome1 & \large Nome2 \tabularnewline
  Department1 & Department2 \tabularnewline
  School1 & School2 \tabularnewline
  \url{url1} & \url{url2}
  \end{tabular}
}

\title{Diagnostics for Semiparametric Accelerated Failure Time Models with  
\proglang{R} Package \pkg{afttest}}
\Plaintitle{\textbf{R} Package \texttt{afttest}}
\Shorttitle{\textbf{R} Package \texttt{afttest}}

\author{
  Woojung Bae \\ University of Florida
  \And Dongrak Choi \\ Duke University
  \And Jun Yan \\ University of Connecticut
  \And Sangwook Kang \\ Yonsei University
  }
\Plainauthor{Woojung Bae, Dongrak Choi, Jun Yan, Sangwook Kang}

\Address{
  Woojung Bae \\
  Department of Statistics \\
  University of Florida \\
  Griffin-Floyd Hall, Gainesville, FL 32611, USA. \\
  E-mail: \email{woojung.bae@ufl.edu}
  
  Dongrak Choi \\
  Department of Biostatistics and Bioinformatics \\
  Duke University \\
  Hock Plaza, 2424 Erwin Rd, Durham, NC 27705, USA. \\
  E-mail: \email{dongrak.francisco.choi@gmail.com}
  
  Jun Yan\\
  Department of Statistics\\
  University of Connecticut\\
  215 Glenbrook Rd. Unit 4120, Storrs, CT 06269, USA.\\
  Email: \email{jun.yan@uconn.edu}
  
  Sangwook Kang \\
  Department of Applied Statistics and Department of Statistics and Data Science \\
  Yonsei University \\
  50 Yonsei-ro, Seodaemun-gu, Seoul 03722, Republic of Korea \\
  E-mail: \email{kanggi1@yonsei.ac.kr} 
}

\Abstract{
The semiparametric accelerated failure time (AFT) model offers a
direct and interpretable alternative to the Cox proportional hazards
model, yet practical diagnostic tools for this framework remain
limited. We introduce \pkg{afttest}, an
\proglang{R} package that implements martingale-residual-based
goodness-of-fit procedures for semiparametric AFT models. In addition
to the recently developed multiplier bootstrap diagnostics, the
package introduces a new computationally efficient resampling strategy
based on an influence-function linear approximation. Unlike the
original approach, which requires repeatedly solving estimating
equations for each bootstrap replicate, the proposed method avoids
iterative optimization and substantially reduces computation time
while preserving asymptotic validity. Both the standard multiplier
bootstrap and the accelerated linear approximation are implemented,
allowing users to balance finite-sample performance and computational
scalability. The package supports rank-based and least-squares
estimators, provides omnibus, link function, and functional form
tests, and includes graphical tools for visualizing residual
processes. An application to the Mayo Clinic primary biliary cirrhosis
study illustrates the workflow.
}
\Plainkeywords{Induced smoothing procedure, Kolmogorov-type test, 
Martingale residual, Rank-based estimation, Resampling, Survival analysis}

\IfFileExists{upquote.sty}{\usepackage{upquote}}{}
\begin{document}
\pagenumbering{arabic}
\section{Introduction} \label{sec:introduction}
Survival analysis, which investigates time-to-event data with censoring, often 
relies on strong assumptions, necessitating the use of robust statistical models. 
The Cox proportional hazard (PH) model and the semiparametric accelerated 
failure time (AFT) model are two of the most prominent approaches.
The Cox PH model, also known as the relative risk model, 
has become a cornerstone in survival analysis due to its flexibility 
and the availability of well-established inference and model-checking
procedures~\citep{cox1972regression, spiekerman1996checking,
  ying1993checking}. Software
implementations such as the \pkg{survival} package~\citep{therneau2023package} 
and the \proglang{SAS} \code{PROC PHREG} procedure~\citep{sas2016} have made 
these tools accessible to medical and public health researchers, offering 
functions for parameter estimation and goodness-of-fit testing. Model 
diagnostics often rely on techniques like the weighted sum of the martingale 
process~\citep{spiekerman1996checking}. Additionally, the
\pkg{timereg} package~\citep{scheike2006dynamic, scheike2011analyzing}
in \proglang{R} extends the
modeling framework by providing flexible options like the Cox–Aalen model, 
with resampling-based methods used for hypothesis testing and $p$-value 
approximation.

The Cox model's dependence on the PH
assumption---that hazard ratios remain constant over time---can be
overly restrictive, as this condition is frequently violated in 
practice. Moreover, its inability to directly estimate the baseline hazard 
function limits absolute risk estimation and can result in biased conclusions 
when the assumption is violated. In contrast, the AFT
model~\citep{buckley1979linear, kalbfleisch2011statistical} provides a
linear framework that directly models
the logarithm of failure time, offering regression parameters 
that are more straightforward to interpret.
The semiparametric AFT model avoids the need to specify a particular error
distribution and typically employs rank-based estimation methods grounded in
the weighted log-rank test~\citep{prentice1978linear}. Extensive theoretical
developments~\citep{tsiatis1990estimating, ying1993checking} further support
these methods, making the AFT model a compelling alternative when the
PH assumption of the Cox model is questionable.

Existing estimation methods for the semiparametric AFT model are well
developed, yet diagnostic methodology remains comparatively
limited. Rank-based estimators facilitated by induced smoothing
\citep{brown2006induced, chiou2015rank} and least-squares procedures
with stable iterative algorithms \citep{jin2006least,
  chiou2014marginal, son2023quantile} have been extensively studied and implemented in
\proglang{R} packages, including \pkg{aftgee}~\citep{chiou2014fitting}.
In contrast, model-checking tools have received far
less attention. Recent work constructed goodness-of-fit procedures
using cumulative sums of martingale residuals~\citep{choi2024general},
extending earlier omnibus testing
ideas~\citep{novak2013goodness}. These developments introduced
omnibus, link function, and covariate functional form tests and were
accompanied by an initial implementation in the companion package
\pkg{afttest}~\citep{bae2022afttest}.

The primary computational challenge in residual-based diagnostics
arises from the multiplier bootstrap required to approximate the null
distribution. The implementation in \citet{choi2024general} required
solving the estimating equations for each bootstrap replicate, which
becomes computationally intensive as the sample size or number of
resampling paths increases. To overcome this limitation, we introduce
a computationally efficient resampling strategy based on an asymptotic
linear approximation of the residual process. By exploiting the
influence-function representation of the estimator, the proposed
method replaces iterative re-estimation with direct evaluation of
perturbed influence terms. This approach preserves asymptotic validity
while substantially reducing computation time, thereby making routine
diagnostic analysis feasible for moderate and large datasets. Both the
original bootstrap and the accelerated approximation are implemented
to allow comparison.

The goal of this article is to document the computational architecture
and practical workflow of the \pkg{afttest} package. The package
integrates with \pkg{aftgee} to establish a coherent
estimation–diagnosis pipeline: models are fitted using \pkg{aftgee},
and adequacy is subsequently assessed using \pkg{afttest}. The
software adopts a unified S3 interface supporting rank-based and
least-squares estimators and provides dedicated \code{plot()} methods
for visualizing stochastic processes underlying the
tests. Optimization is carried out using the DF-SANE
algorithm~\citep{varadhan2009bb}. Computational efficiency is enhanced
through \pkg{Rcpp} and
\pkg{RcppArmadillo}~\citep{eddelbuettel2011rcpp,
  eddelbuettel2014rcpparmadillo}, while formula handling and
visualization rely on \pkg{survival}~\citep{therneau2023package},
\pkg{ggplot2}~\citep{hadley2009ggplot2}, and
\pkg{gridExtra}~\citep{auguie2017gridExtra}.

The remainder of the article is organized to distinguish the 
methodological development from its software implementation and 
empirical evaluation. Section~\ref{sec:test} reviews the residual-based 
test statistics introduced in \citet{choi2024general} and presents the 
new computationally efficient linear-approximation strategy for 
resampling, together with its theoretical justification. A simulation 
study in Section~\ref{sec:test} compares the accelerated approach with 
the original multiplier bootstrap reported in \citet{choi2024general},
evaluating finite-sample performance and computational
efficiency. Section~\ref{sec:usage} documents the architecture and 
usage of the \proglang{R} package, including supported estimators and 
available function options. Section~\ref{sec:illustration} applies the 
procedures to the primary biliary cirrhosis (PBC) study data. 
Section~\ref{sec:discussion} concludes with discussion and potential 
extensions.

\section{Test Statistics and Computational Strategy} \label{sec:test}

The diagnostic procedures in the \pkg{afttest} package are based on 
multi-parameter stochastic processes constructed from martingale residuals. This 
section outlines the notation, the computational bottleneck of the standard 
approach, and the efficient linear approximation strategy implemented in the package.

\subsection{Basis Stochastic Process} \label{sec:test.basis}

Consider a sample of $n$ independent subjects. For
the~$i$th subject, $i = 1, \ldots, n$, let $T_{i}$ denote the potential
failure time and $C_{i}$ the censoring time. The observed data are
$\left( X_{i}, \Delta_{i}, \bZ_{i} \right)$, $i = 1, \ldots, n$,
where $X_{i} = \min \left\{ T_{i}, C_{i} \right\}$, $\Delta_{i} = I(T_{i} \le C_{i})$, 
$I ( \cdot )$ is the indicator function, and 
$\bZ_{i} = \left( Z_{i1}, \ldots, Z_{ip} \right)^{\top}$ is a bounded
$p$-dimensional covariate vector.
Given $\bZ_{i}$, we assume that $T_{i}$ and $C_{i}$ are 
independent for all subjects. The semiparametric AFT model specifies
that
\begin{align} \label{eq:test:1}
    \log T_{i} = - \bZ_{i}^{\top} \bb_{0} + \epsilon_{i},
\end{align}
where $\bb_{0} = \left( \beta_{01}, \ldots, \beta_{0p} \right)^{\top}$ is a 
vector of true regression parameters and $\epsilon_{i}$ is an unspecified random 
error term with mean zero.

To check the adequacy of this model, \citet{choi2024general} presented
multiple statistics based on a class of multi-parameter stochastic
processes. Let $\Mh[t; \bbh[n] ][n,i]$ denote the estimated martingale
residual for the $i$th subject at time $t$, where $\bbh[n]$ is a
$\sqrt{n}$-consistent estimator of $\bb_{0}$ (e.g., the rank-based or
least-squares estimator provided by package \pkg{aftgee}). We define the
observed process $W_{n}(t, \bz; \bbh[n])$ as:
\begin{align} \label{eq:test:2}
    \bW[t, \bz ; \bbh[n]][n] = n^{- \frac{1}{2}} \sum_{i=1}^{n} \pi_{i} 
    \left( \bz \right) \Mh[t; \bbh[n] ][n,i],
\end{align}
where $\pi_{i} \left( \bz \right) = \ell \left( \bz; \bZ_{i} \right) I 
\left( \bZ_{i} \leq \bz \right)$ with a bounded function of $\bz$, $\ell 
\left( \cdot \right)$ and the choice of $\pi_{i} \left( \bz \right)$ determines 
the specific type of model diagnostic as detailed in Section \ref{sec:test.specific}.

To assess the significance of testing statistics that are functionals
of the observed basis stochastic process $\bW[t, \bz ; \bbh[n]][n]$,
one need to approximate the null distribution of
$\bW[t, \bz ; \bbh[n]][n]$.
\citet{choi2024general} proposed a perturbed process,
denoted here as $\bWh[t, \bz; \bbh[n]][n]$, based on a multiplier
bootstrap procedure.  This approach is
computationally expensive because it requires solving complex estimating 
equations via numerical optimization for every bootstrap iteration. In contrast, 
the efficient closed-form calculation introduced in the next section avoids this 
repetitive optimization.

\subsection{Efficient Resampling of the Basis Stochastic Process} \label{sec:test.efficient}

To address the computational burden of iterative resampling, the 
\pkg{afttest} package implements an asymptotically equivalent statistic, 
denoted by $\bWh[t, \bz; \bbh[n]][n][\dag]$, that avoids repeated 
re-estimation of model parameters. The construction relies on the 
influence-function representation of the residual process, which admits 
the expansion 
$\bW[t, \bz ; \bbh[n]][n] = \sum_{i=1}^n h_{i}(t, \bz; \bb_0) + o_p(1)$, 
where $h_{i}(t, \bz; \bb_0)$ denotes the $i$th influence function. 
This representation was established in \citet{choi2024general}, 
with the explicit form of $h_{i}(\cdot)$ given in Appendix~A.1 therein. 
Let $\phi_{i}$ ($i=1, \ldots, n$) be independent multiplier random 
variables with $\E[\phi_{i}] = \VAR(\phi_{i}) = 1$. The proposed perturbed 
process is defined as
\begin{align} \label{sec:test:3}
    \bWh[t, \bz; \bbh[n]][n][\dag] 
    = n^{-1/2} \sum_{i=1}^n (\phi_{i} - 1) 
    \hat{h}_{i} \left( t, \bz; \bbh[n] \right),
\end{align}
where $\hat{h}_{i}(t, \bz; \bbh[n])$ denotes the estimated $i$th influence 
function. Appendix~\ref{sec:multiplier} establishes that 
$\bWh[t, \bz; \bbh[n]][n][\dag]$ shares the same limiting distribution as 
$\bWh[t, \bz; \bbh[n]][n]$.

The bootstrap procedure implemented in \pkg{afttest} proceeds as follows:
\begin{enumerate}
    \item Fit the model once to obtain $\bbh[n]$. 
    
    \item Compute the observed test statistic $\mathcal{W}_{\nobs} = \sup \abs{\bW[t, \bz; \bbh[n]][n]}$.
    
    \item For $b = 1, \dots, B$ (where $B$ is \code{npath}):
    \begin{enumerate}
        \item Generate independent and identically distributed (i.i.d.) positive 
        multiplier random weights $\left( \phi_{1}, \cdots, \phi_{n} \right)$ 
        from a distribution with mean 1 and variance 1, e.g., $\phi_{i} \sim \text{exp}(1)$.

        \item Compute the fast perturbed process $\bWh[t, \bz; \bbh[n]][n][\dag]$ 
        using Equation \eqref{sec:test:3}. This approach bypasses the computationally 
        intensive step of obtaining a perturbed estimator $\bbh[n][\phi]$ for every 
        replicate by iteratively solving the estimating equations 
        $\score[\infty, \bb][n] = \score[\infty, \bbh[n]][n][\phi]$.
        
        \item Compute the supremum statistic $\mathcal{W}_{\napp}^{(b)} = 
        \sup \abs{\bWh[t, \bz; \bbh[n]][n][\dag]}$.
        
    \end{enumerate}
    
    \item Calculate the $p$-value as $B^{-1} \sum_{b=1}^{B} I \left( 
    \mathcal{W}_{\napp}^{(b)} \ge \mathcal{W}_{\nobs} \right)$.
    
\end{enumerate}

The repeated optimization required by the resampling scheme in 
\citet{choi2024general} (Step~3(b)) imposes substantial computational 
cost, particularly when the sample size or the number of resampling 
paths is large \citep{kojadinovic2012goodness}. The linear-approximation 
strategy implemented here eliminates this repeated re-estimation while 
preserving the asymptotic distribution of the statistic.

To improve finite-sample stability and facilitate interpretation, a 
standardized version of the supremum statistic is also available. The 
algorithm described above computes the unstandardized statistic based 
on the perturbed process. The standardized statistic rescales this same 
process by an estimated variance function and is defined as
\[
\sup \left| \bWh[t, \bz; \bbh[n]][n][\dag] 
        / \hat{\sigma}_{n}(t, \bz) \right|,
\]
where $\hat{\sigma}_{n}^2(t, \bz)$ is estimated from the empirical 
variance of the generated resampling paths, following 
\citet{choi2024general}.

\subsection{Specific Tests} \label{sec:test.specific}

To evaluate how well the data fit the AFT model, we consider three
types of goodness-of-fit tests: an omnibus test for overall model fit,
a test for the link function, and a test for the functional form of
individual covariates.

The omnibus test checks for general departures from the model by examining both 
time and covariates. The corresponding null hypotheses are $\text{H}_{0}$: The 
assumed semiparametric AFT model $\log T_{i} = - \bZ_{i}^{\top} \bb_{0} + 
\epsilon_{i}$ fits the observed data adequately. vs. $\text{H}_{1}$: any 
departure from $\text{H}_{0}$. The test statistic is defined as: 
\begin{align*} 
    \sup_{t, \bz} \abs{\bW[t, \bz; \bbh[n]][n]} \quad \text{where} \quad 
    \bW[t, \bz; \bbh[n]][n] = n^{- \frac{1}{2}} \sum_{i=1}^{n} I \left( \bZ_{i} 
    \leq \bz 
    \right) \Mh[t; \bbh[n]][n,i]. 
\end{align*}

The link function test evaluates whether the relationship between covariates and 
the log survival time is correctly specified. In the general case, the model is 
specified as: 
\begin{align*} 
    \log T_{i} &= g \left( \bZ_{i}^{\top} \bb \right) + \epsilon_{i}, 
\end{align*} 
where $g \left( \cdot \right)$ is an unknown function. If the model is correct, 
$g \left( \cdot \right)$ is an identity function, \textit{i.e.}, $\text{H}_{0}$: $g (x) 
= x$ vs. $\text{H}_{1}$: $g (x) \ne x$ for some $x$. The test statistic is: 
\begin{align*} 
    \sup_{\bz} \abs{\bW[\bz][n]} \quad \text{where} \quad \bW[\bz][n] = n^{- 
    \frac{1}{2}} \sum_{i=1}^{n} I \left( \bZ_{i} \leq \bz \right) \Mh[\infty; 
    \bbh[n]][n,i]. 
\end{align*} 
This is a special case of the omnibus test where $t = \infty$.

Lastly, the functional form test checks if each covariate enters the model 
linearly. A general form of the AFT model is: 
\begin{align*} 
    \log T_{i} &= g \left( \bZ_{i} \right)^{\top} \bb + \epsilon_{i}, 
\end{align*} 
where $g \left( \bZ_{i} \right) = \left\{ g_{1} (Z_{i1}), \ldots, g_{p} (Z_{ip}) 
\right\}$ and each $g_{q} \left( Z_{iq} \right), q = 1, \ldots, p$ allows 
for a flexible transformation of the $q^{\text{th}}$ covariate. If the model 
holds, each $g_{q} \left( \cdot \right)$ is an identity function, \textit{i.e.}, 
$\text{H}_{0}$: $g_{q} (x) = x$ vs. $\text{H}_{1}$: $g_{q} (x) \ne x$ for some 
$x$. The test statistic for the $q^{\text{th}}$ covariate is: 
\begin{align*} 
    \sup_{\bz} \abs{ \bW[\bz][n, q] } \quad \text{where} \quad 
    \bW[\bz][n, q] = n^{- \frac{1}{2}} \sum_{i=1}^{n} I \left( Z_{iq} \leq z_{q} 
    \right) 
    \Mh[\infty ; \bbh[n]][n,i]. 
\end{align*}
This constitutes a special case of the link function test in which 
$z_{r} = \infty$ for all $r \ne q$.


The proposed procedure is applicable to any estimators for $\bb_{0}$ that satisfy 
the conditions mentioned above; see~\citet{choi2024general}.

\subsection{Simulation Study} \label{sec:test.simulation}

To evaluate the finite-sample performance of the proposed method, we conducted 
simulation experiments. For simulation setups, we adopted the same simulation 
setup as presented in~\citet{choi2024general}. However, the current analysis 
introduces two key distinctions: first, we additionally implemented the least 
squares method alongside the induced smoothed and non-smoothed methods examined 
in the previous study; second, a linear approximation strategy was implemented 
to expedite computation. Detailed descriptions of the data generating process 
and the full rejection rate results (Table~\ref{tab:sim:result}) are deferred to 
Section~\ref{sec:detailed}.

The simulation results demonstrate that the proposed linear approximation 
strategy yields Type \Romannum{1} error rates and statistical power comparable 
to the method without linear approximation proposed by \citet{choi2024general}. 
While their original approach exhibits slightly higher power at smaller sample 
sizes (\textit{e.g.}, $n=100$), the performance of the two methods becomes 
nearly identical as the sample size increases to $n=500$, as detailed in 
Table~\ref{tab:sim:result}. Consequently, for large sample sizes, the proposed 
method maintains equivalent statistical validity while providing a distinct 
advantage in computational efficiency.

While running time is contingent on factors such as sample size, covariate count, 
and data complexity, Table~\ref{tab:TIMEsim:result} summarizes the specific 
timing results for different sample sizes and values of $\gamma$. The results 
demonstrate a substantial improvement in computational efficiency using the
proposed asymptotic linear approximation (\code{linApprox = TRUE}). Across all 
scenarios, the proposed method reduces computation time by orders of magnitude 
compared to the standard resampling method (\code{linApprox = FALSE}). For example, 
with $n=500$ and $\gamma=0$, the average time for the omnibus test using the 
non-smooth (ns) estimator drops from 435.9 seconds to just 12.9 seconds. Even 
for the least squares (ls) estimator, which is computationally simpler, the 
proposed method reduces the running time by over 96\%. 
{
\begin{table}[htp]
    \caption{\label{tab:TIMEsim:result} \label{tab:TIMEsim:result} Average 
    running times (in seconds) for the omnibus, link function, and functional 
    form tests based on 1,000 simulations under the Model \eqref{model:sim1}. 
    The table compares the computational efficiency of the proposed asymptotic 
    linear approximation (\code{linApprox = TRUE}) against the standard resampling 
    method (\code{linApprox = FALSE}) for non-smooth (ns), induced-smoothed (is), and 
    least squares (ls) estimators across sample sizes $n \in \{100, 500\}$ with 
    a 20\% censoring rate. Computations were performed using parallel computing 
    on the University of Florida HiPerGator AI cluster (70,320 cores, 18GB RAM 
    per core).}
    \centering
    \begin{tabular}[t]{ccccccccc}
        \toprule
        \multicolumn{3}{c}{ } & \multicolumn{3}{c}{$n = 100$} & \multicolumn{3}{c}{$n = 500$} \\
        \cmidrule(l{3pt}r{3pt}){4-6} \cmidrule(l{3pt}r{3pt}){7-9}
        $\gamma$ & Test & linApprox & ns & is & ls & ns & is & ls\\
        \midrule
        0 & omni & T & 0.8 & 0.3 & 0.5 & 12.9 & 11.3 & 9.4\\
         &  & F & 45.6 & 45.7 & 4.7 & 435.9 & 448.8 & 265.7\\
         & link & T & 0.2 & 0.2 & 0.2 & 6.9 & 6.9 & 5.1\\
         &  & F & 43.5 & 43.6 & 4.1 & 434.4 & 438.9 & 241.1\\
         & form & T & 0.2 & 0.2 & 0.2 & 6.9 & 6.9 & 5.0\\
         &  & F & 43.6 & 43.7 & 4.2 & 438.9 & 440.0 & 225.3\\
        \cmidrule{1-9}
        0.5 & omni & T & 0.3 & 0.3 & 0.3 & 12.8 & 11.2 & 9.4\\
         &  & F & 34.0 & 34.0 & 5.0 & 368.7 & 377.4 & 259.6\\
         & link & T & 0.2 & 0.2 & 0.2 & 6.9 & 6.9 & 5.2\\
         &  & F & 34.1 & 34.1 & 4.8 & 361.3 & 357.6 & 235.0\\
         & form & T & 0.2 & 0.2 & 0.2 & 6.9 & 6.8 & 5.0\\
         &  & F & 34.1 & 34.1 & 4.8 & 351.7 & 335.6 & 192.4\\
        \bottomrule
    \end{tabular}
\end{table}
}

\section{Usage} \label{sec:usage}
The package \pkg{afttest} offers a unified interface to the methods discussed in 
Section~\ref{sec:test} and \citet{choi2024general} via the function 
\code{afttest()}. The function is 
designed to accommodate two common workflows: users may either supply a model 
formula and allow \pkg{afttest} to fit the model internally, or provide a 
pre-fitted model object obtained from the \pkg{aftgee} package. The arguments 
for the generic function and its method-specific implementations are displayed 
below:
\begin{knitrout}
\definecolor{shadecolor}{rgb}{0.969, 0.969, 0.969}\color{fgcolor}\begin{kframe}
\begin{alltt}
\hlkwd{library}\hldef{(afttest)}
\hlkwd{args}\hldef{(}\hlkwd{getS3method}\hldef{(}\hlsng{"afttest"}\hldef{,} \hlsng{"formula"}\hldef{))}
\end{alltt}
\begin{verbatim}
## function (object, data, npath = 200, testType = "omnibus", estMethod = "rr", 
##     eqType = "ns", covTested = 1, npathsave = 50, linApprox = TRUE, 
##     seed = NULL, ...) 
## NULL
\end{verbatim}
\begin{alltt}
\hlkwd{args}\hldef{(}\hlkwd{getS3method}\hldef{(}\hlsng{"afttest"}\hldef{,} \hlsng{"aftsrr"}\hldef{))}  \hlcom{# object from rank-based fitting}
\end{alltt}
\begin{verbatim}
## function (object, data, npath = 200, testType = "omnibus", eqType = "ns", 
##     covTested = 1, npathsave = 50, linApprox = TRUE, seed = NULL, 
##     ...) 
## NULL
\end{verbatim}
\begin{alltt}
\hlkwd{args}\hldef{(}\hlkwd{getS3method}\hldef{(}\hlsng{"afttest"}\hldef{,} \hlsng{"aftgee"}\hldef{))}  \hlcom{# object from least-squares fitting}
\end{alltt}
\begin{verbatim}
## function (object, data, npath = 200, testType = "omnibus", eqType = "ls", 
##     covTested = 1, npathsave = 50, linApprox = TRUE, seed = NULL, 
##     ...) 
## NULL
\end{verbatim}
\end{kframe}
\end{knitrout}

The primary argument, \code{object}, determines the model to be tested. If 
\code{object} is a formula, the function fits the semiparametric AFT model 
using the variables supplied in the optional \code{data} argument; the formula 
must contain a \code{Surv} object from the \pkg{survival} package. If 
\code{object} is instead a fitted model returned by \code{aftsrr()}
(rank-based) or  \code{aftgee()} (least squares) from the \pkg{aftgee}
package, \code{afttest()} performs 
diagnostics directly on that object. Internally, the function uses S3 method 
dispatch to distinguish between formula input and fitted model objects, 
thereby maintaining a consistent user interface while preserving compatibility 
with \pkg{aftgee}.

When \code{object} is a formula, the estimation method is controlled by 
\code{estMethod} and \code{eqType}. The \code{estMethod} argument specifies the 
estimator type from the \pkg{aftgee} package: \code{"rr"} (default) selects the 
rank-based regression method via \code{aftsrr()}, while \code{"ls"} selects the 
least-squares method via \code{aftgee()}. The \code{eqType} argument is specific 
to rank-based estimation (\code{estMethod = "rr"}) and determines the estimating 
equation: \code{"ns"} uses the non-smoothed Gehan-based equation, while 
\code{"is"} uses the induced-smoothed version. For further details on these 
estimators, refer to the \pkg{aftgee} package documentation~\citep{chiou2014fitting}.

To control the scope and precision of the inference, the \code{testType} argument 
specifies the type of test to be conducted: the omnibus test 
(\code{testType = "omnibus"}), the link function test (\code{testType = "link"}), 
or the functional form test (\code{testType = "covForm"}). The default is the 
omnibus test. The \code{npath} argument determines the number of approximated 
processes used to estimate the null distribution, with a default of 200. To 
ensure stability, the minimum value for \code{npath} is set to 10; values less 
than 10 are automatically adjusted. To ensure the reproducibility of these 
resampling results, the optional \code{seed} argument accepts an integer to 
initialize the random number generator.

Specific arguments accommodate different testing needs. The \code{covTested} 
argument is applicable only when \code{testType = "covForm"} and specifies the 
continuous covariate to be tested. Users can specify this covariate by its 
numeric index (defaulting to \code{1}) or its character name. Note that 
functional form tests are not performed for binary or categorical variables, as 
discrete covariates inherently satisfy the linearity assumption.

To address computational efficiency, the \code{linApprox} argument controls the 
resampling technique. If \code{TRUE} (default), the function employs an 
asymptotic linear approximation to significantly speed up bootstrap computations; 
if \code{FALSE}, it utilizes the standard multiplier bootstrap method described 
in \citet{choi2024general}. Additionally, the \code{npathsave} argument 
(defaulting to 50) allows users to specify the number of perturbed paths to save. 
Users should exercise caution with this parameter, as storing a large number of 
paths can be memory-intensive. For example, in the omnibus test, each path 
requires storing an $n \times n$ matrix, leading to a total storage requirement 
of $\text{\code{npath}} \times n^2$ elements.

The function returns a list object inheriting from classes \code{afttest} and 
\code{htest}. This object contains the estimated regression coefficients 
(\code{beta}), the estimated standard error of the observed process 
(\code{SE_process}), the observed process (\code{obs_process}), the approximated 
processes (\code{apprx_process}), and their standardized counterparts 
(\code{obs_std_process} and \code{apprx_std_process}). Two sets of $p$-values 
are provided—unstandardized (\code{p_value}) and standardized 
(\code{p_std_value})—though the standardized version is 
recommended~\citep{choi2024general}. The output also preserves input arguments 
such as \code{npath}, \code{eqType}, \code{testType}, and \code{estMethod}, 
along with a data frame of the observed failure times, failure indicators, and 
scaled covariates. Consistent with the memory considerations for 
\code{npathsave}, the returned \code{obs_process} and \code{apprx_process} for 
the omnibus test are stored as $n \times n$ matrices (ordered by weight rank vs. 
time-transformed residual rank), whereas the link function and functional form 
processes are stored as more compact $n \times 1$ vectors.

The S3 method to plot a \code{afttest()} object is shown below.
\begin{knitrout}
\definecolor{shadecolor}{rgb}{0.969, 0.969, 0.969}\color{fgcolor}\begin{kframe}
\begin{alltt}
\hlkwd{plot}\hldef{(x,} \hlkwc{npath} \hldef{=} \hlnum{50}\hldef{,} \hlkwc{std} \hldef{=} \hlnum{TRUE}\hldef{,} \hlkwc{quantile} \hldef{=} \hlkwa{NULL}\hldef{)}
\end{alltt}
\end{kframe}
\end{knitrout}
The \code{plot()} function requires only one argument, \code{x}, which
is the result obtained from the \code{afttest()} function. The function
automatically generates a graph corresponding to the \code{testType} of the
result. If the argument is incorrectly specified, the function will return an
error message \code{"Must be afttest class"}. The \code{npath} argument is used
to specify the number of simulation paths to be plotted in the graph and should
not exceed the number used in the \code{afttest()} function. If \code{npath} is
greater than the number used in \code{afttest()}, it will be set to the same
value. The default value for \code{npath} is \code{50}. The \code{std}
argument determines whether to plot the standardized process 
(\code{std = TRUE}) or unstandardized process 
(\code{std = FALSE}), with a default value of \code{std = TRUE}. 
The graph is generated using the \pkg{ggplot2}
~\citep{hadley2009ggplot2} and \pkg{gridExtra}~\citep{auguie2017gridExtra}
packages. The argument \code{quantile} allows users to specify five quantiles 
corresponding to the quantiles of $\bz$ in $\pi_{i} \left( \bz \right) = 
I \left( \bZ_{i} \leq \bz \right)$, with default values set to 10\%, 25\%, 
50\%, 75\%, and 90\%. This argument is applicable only when \code{testType = 
"omnibus"}. Since the omnibus test produces the form of 
$\text{\code{npath}} \times n \times n$ matrices, selecting a subset of 
$\bz$ values is necessary to generate two-dimensional plots. For the Cox 
PH model, similar goodness-of-fit diagnostics based on simulating sample paths 
from residual-based stochastic processes are already well-established including 
the \proglang{SAS} \code{PROC PHREG} procedure's \code{ASSESS} 
statement~\citep{sas2016}

\section{Illustration} \label{sec:illustration}
For a practical demonstration of the proposed method, we analyze the PBC 
data from the Mayo clinic~\citep{fleming2013counting}. 
PBC is an autoimmune liver disease that is characterized by progressive destruction 
of the intrahepatic bile ducts. The PBC dataset comprises a comprehensive collection 
of clinical and laboratory data on 418 patients who were treated for PBC at the Mayo 
clinic between 1974 and 1984. In a previous study, \citet{dickson1997primary} 
utilized the PBC dataset to develop a Cox PH model of the 
medical history of the disease, incorporating six covariates:
\code{bili} (bilirubin), a measure of serum bilirubin where elevated
levels indicate the liver's reduced ability to excrete waste products;
\code{protime}, the time it takes for blood to clot, with longer times
reflecting the liver's impaired ability to synthesize clotting
factors; \code{albumin}, the level of albumin protein in the blood,
serving as an indicator of the liver's synthetic function; \code{age}, the
patient's age at enrollment; and \code{edema}, a clinical measure of fluid
accumulation representing the severity of liver decompensation.

We first transformed several key variables to analyze the PBC dataset
and evaluate the suitability of alternative functional forms of the
\code{bili}. The failure indicator was
similarly transformed, with a value of \code{1} representing an event
and \code{0} representing censoring.

\begin{knitrout}
\definecolor{shadecolor}{rgb}{0.969, 0.969, 0.969}\color{fgcolor}\begin{kframe}
\begin{alltt}
\hldef{pbc} \hlkwb{<-} \hlkwd{within}\hldef{(survival}\hlopt{::}\hldef{pbc, \{}
    \hldef{status} \hlkwb{<-} \hlkwd{ifelse}\hldef{(status} \hlopt{==} \hlnum{2}\hldef{,} \hlnum{1}\hldef{,} \hlnum{0}\hldef{)}
    \hldef{log_bili} \hlkwb{<-} \hlkwd{log}\hldef{(bili)\})}
\end{alltt}
\end{kframe}
\end{knitrout}

Two models are considered for illustration:
\begin{gather}
  \log T_{i} = - \mathtt{bili}_{i} \beta_{1} - \mathtt{protime}_{i} \beta_{2} - \mathtt{albumin}_{i} \beta_{3} - \mathtt{age}_{i} \beta_{4} - \mathtt{edema}_{i} \beta_{5} + \epsilon_{i}, \tag{M1} \label{eq:illustration:1} \\
  \log T_{i} = - \mathtt{log\_bili}_{i} \beta_{1} - \mathtt{protime}_{i} \beta_{2} - \mathtt{albumin}_{i} \beta_{3} - \mathtt{age}_{i} \beta_{4} - \mathtt{edema}_{i} \beta_{5} + \epsilon_{i}. \tag{M2} \label{eq:illustration:2} 
\end{gather}
Model~\ref{eq:illustration:1} includes
covariates \code{bili}, \code{protime}, \code{albumin}, \code{age}, and
\code{edema} without any transformation, while
Model~\ref{eq:illustration:2} modifies \eqref{eq:illustration:1}
by replacing \code{bili} with \code{log_bili}.

\subsection{Model~M1} \label{sec:illustration1}

The analysis can be performed using two different approaches depending
on the specific test being run. For the \code{testType = "omnibus"}
and \code{testType = "link"} tests, we first created a model object
using \code{aftsrr()} in package \pkg{aftgee}. This fitted object was
then passed to \code{afttest()}, which automatically dispatched the
\code{afttest.aftsrr} method to perform the diagnostics. For the
functional form test for \code{bili}, we passed the model formula
directly to \code{afttest()}. This approach automatically uses the
\code{afttest.formula} method, which handles both model fitting and
testing in a single step. For these analyses, the estimation method
was set to \code{estMethod = "rr"}, which uses the \code{aftsrr()}
function from the \proglang{R} package
\pkg{aftgee}~\citep{chiou2014fitting}. When using the rank-based
regression method—either by providing a pre-fitted \code{aftsrr()}
object or by using a formula with \code{estMethod = "rr"}—user can
specify the type of estimating equation with the \code{eqType}
argument. The two available options are \code{"ns"} for the
non-smoothed method and \code{"is"} for the induced-smoothed
method. For all diagnostics, we applied the non-smoothed method for
the omnibus test (\code{eqType = "is"}). One can specify \code{eqType
  = "is"} to use the induced-smoothed method. Finally, the number of
approximated processes was set to 200 using \code{npath = 200}.

First, we fitted the semiparametric AFT model using \code{aftgee::aftsrr()} and 
then plugged in \code{aftsrr()} object to the \code{afttest()} directly. 
\begin{knitrout}
\definecolor{shadecolor}{rgb}{0.969, 0.969, 0.969}\color{fgcolor}\begin{kframe}
\begin{alltt}
\hldef{pbc1_aftsrr} \hlkwb{<-} \hldef{aftgee}\hlopt{::}\hlkwd{aftsrr}\hldef{(}\hlkwd{Surv}\hldef{(time, status)} \hlopt{~} \hldef{bili} \hlopt{+} \hldef{protime} \hlopt{+} \hldef{albumin} \hlopt{+}
                        \hldef{age} \hlopt{+} \hldef{edema,} \hlkwc{data} \hldef{= pbc,} \hlkwc{eqType} \hldef{=} \hlsng{"ns"}\hldef{,}
                      \hlkwc{rankWeights} \hldef{=} \hlsng{"gehan"}\hldef{)}
\end{alltt}
\end{kframe}
\end{knitrout}

\begin{knitrout}
\definecolor{shadecolor}{rgb}{0.969, 0.969, 0.969}\color{fgcolor}\begin{kframe}
\begin{alltt}
\hlkwd{system.time}\hldef{(}
  \hldef{pbc1_omni_ns} \hlkwb{<-}
    \hlkwd{afttest}\hldef{(pbc1_aftsrr,} \hlkwc{data} \hldef{= pbc,} \hlkwc{npath} \hldef{=} \hlnum{200}\hldef{,} \hlkwc{testType} \hldef{=} \hlsng{"omnibus"}\hldef{,}
            \hlkwc{estMethod} \hldef{=} \hlsng{"rr"}\hldef{,} \hlkwc{eqType} \hldef{=} \hlsng{"ns"}\hldef{,} \hlkwc{linApprox} \hldef{=} \hlnum{TRUE}\hldef{,} \hlkwc{seed} \hldef{=} \hlnum{0}\hldef{))}
\end{alltt}
\begin{verbatim}
##    user  system elapsed 
##   1.206   0.075   1.323
\end{verbatim}
\begin{alltt}
\hldef{pbc1_omni_ns} \hlcom{# equivalent to print(pbc1_omni_ns).}
\end{alltt}
\begin{verbatim}
## 
## 	AFT Goodness-of-Fit Test
## 
## Call: 
## afttest.aftsrr(object = pbc1_aftsrr, data = pbc, npath = 200, 
##     testType = "omnibus", eqType = "ns", linApprox = TRUE, seed = 0, 
##     estMethod = "rr")
## 
## --- Null Hypothesis (H0) ---
## The assumed semiparametric AFT model fits the data adequately. 
## 
## --- p-values ---
##  unstandardized standardized
##           0.150        0.005
\end{verbatim}
\end{kframe}
\end{knitrout}

\begin{knitrout}
\definecolor{shadecolor}{rgb}{0.969, 0.969, 0.969}\color{fgcolor}\begin{kframe}
\begin{alltt}
\hlkwd{plot}\hldef{(pbc1_omni_ns,} \hlkwc{std} \hldef{=} \hlnum{TRUE}\hldef{)}
\end{alltt}
\end{kframe}
\end{knitrout}

%
%

\begin{figure}[tbp]
  \centering
  \begin{subfigure}{0.49\textwidth}
    \centering
    \captionsetup{justification=raggedright,singlelinecheck = true}
    \includegraphics[width=0.99\textwidth]{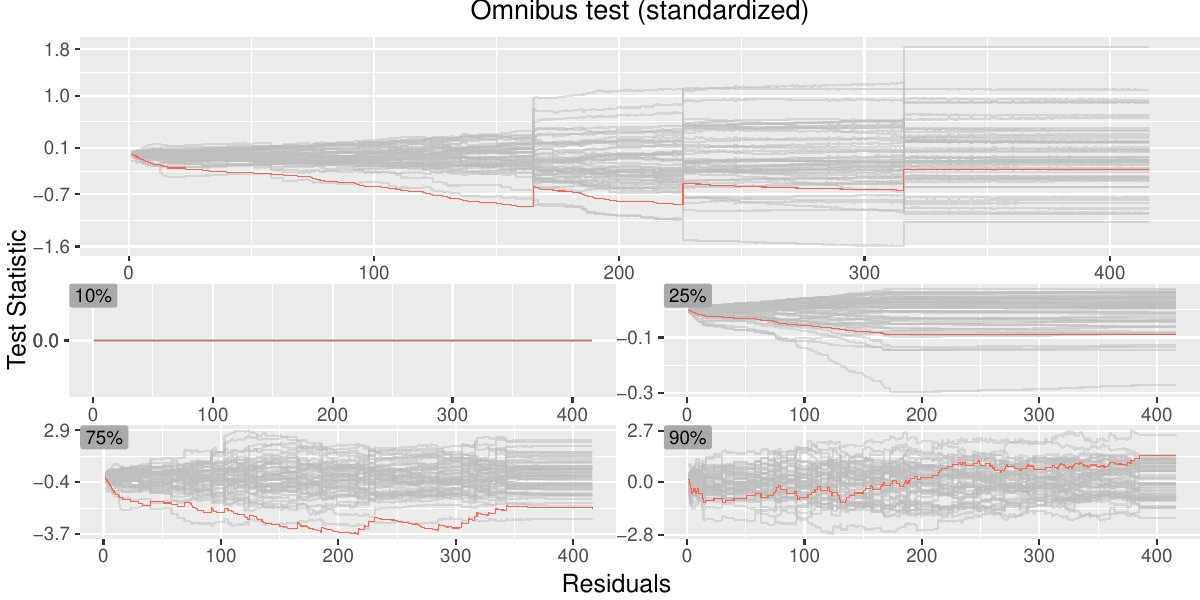}
    \caption{omnibus test} \label{fig:pbc01_omni_ns_std}
  \end{subfigure}
  \begin{subfigure}{0.49\textwidth}
    \centering
    \captionsetup{justification=raggedright,singlelinecheck = true}
    \includegraphics[width=0.99\textwidth]{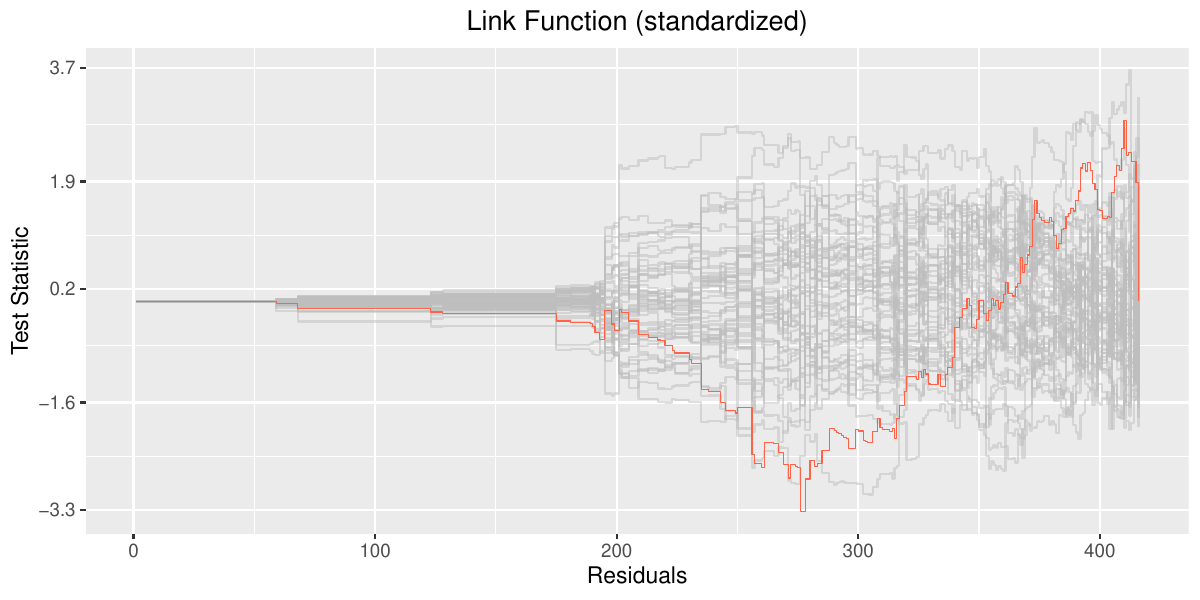}
    \caption{link function test} \label{fig:pbc01_link_ns_std}
  \end{subfigure}
  \\
  \begin{subfigure}{0.49\textwidth}
    \centering
    \captionsetup{justification=raggedright,singlelinecheck = true}
    \includegraphics[width=0.99\textwidth]{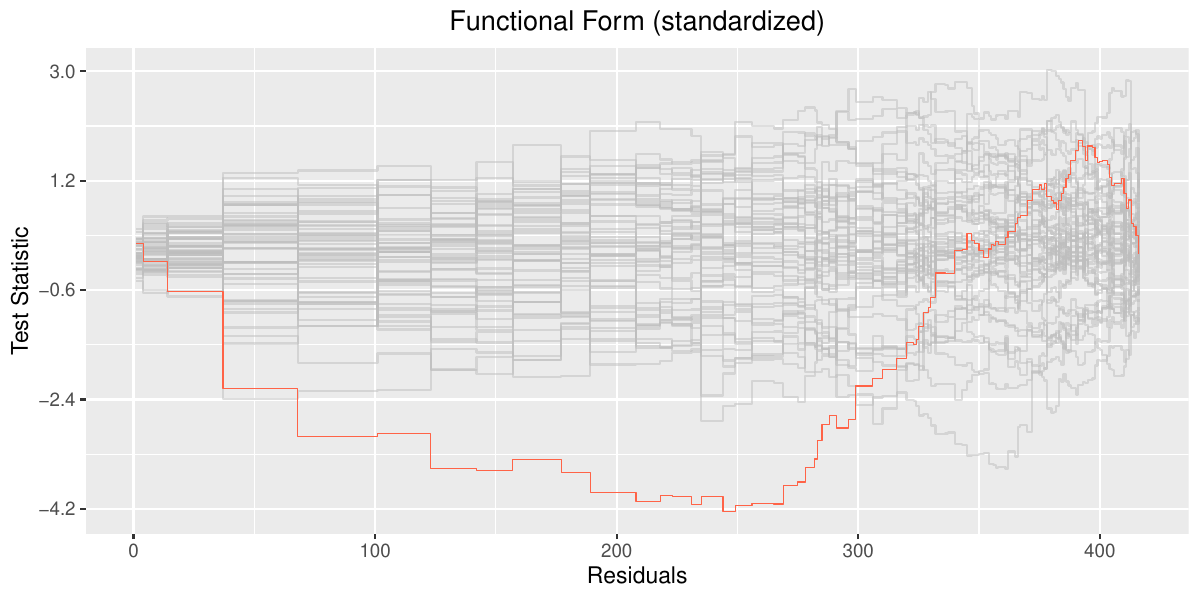}
    \caption{functional form test (\texttt{bili})} \label{fig:pbc01_form1_ns_std}
  \end{subfigure}
  \begin{subfigure}{0.49\textwidth}
    \centering
    \captionsetup{justification=raggedright,singlelinecheck = true}
    \includegraphics[width=0.99\textwidth]{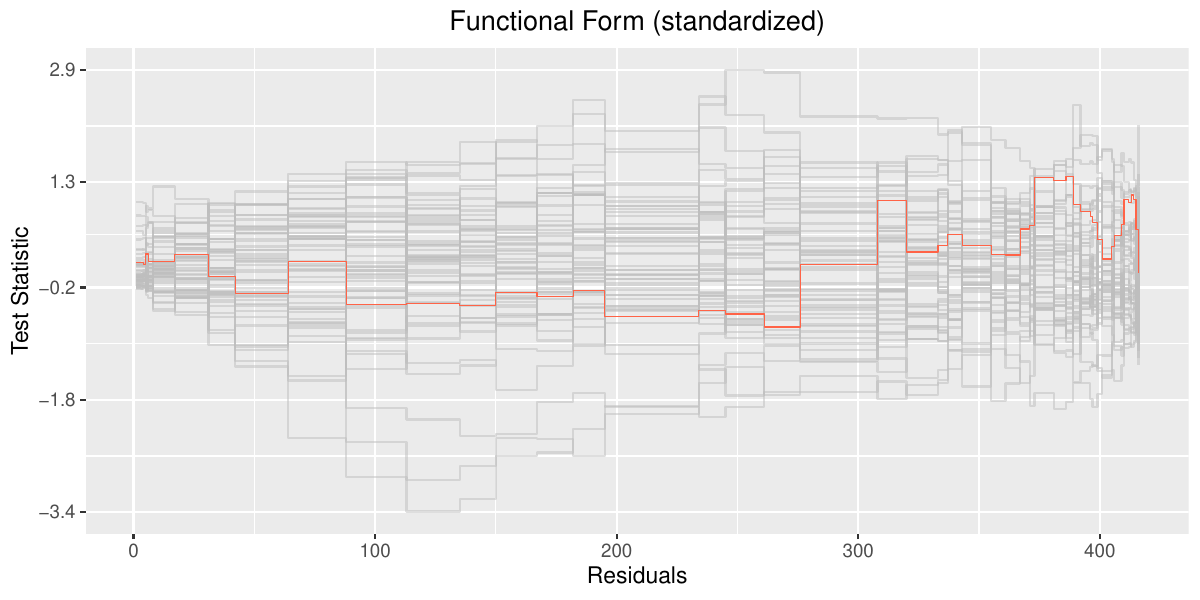}
    \caption{functional form test (\texttt{protime})} \label{fig:pbc01_form2_ns_std}
  \end{subfigure}
  \\
  \begin{subfigure}{0.49\textwidth}
    \centering
    \captionsetup{justification=raggedright,singlelinecheck = true}
    \includegraphics[width=0.99\textwidth]{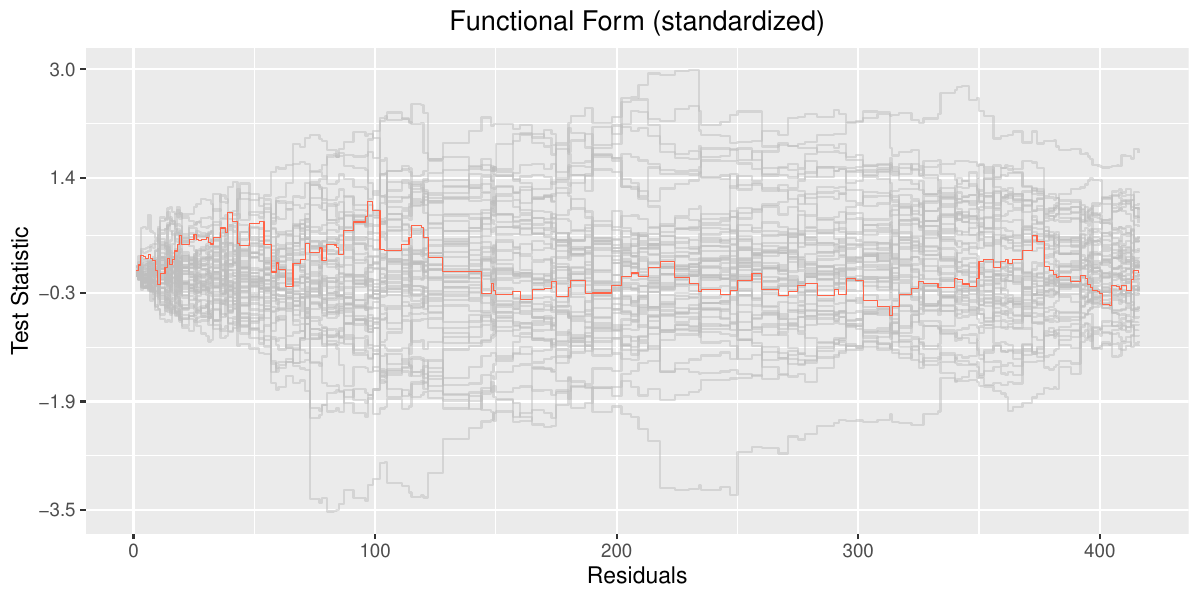}
    \caption{functional form test (\texttt{albumin})} \label{fig:pbc01_form3_ns_std}
  \end{subfigure}
  \begin{subfigure}{0.49\textwidth}
    \centering
    \captionsetup{justification=raggedright,singlelinecheck = true}
    \includegraphics[width=0.99\textwidth]{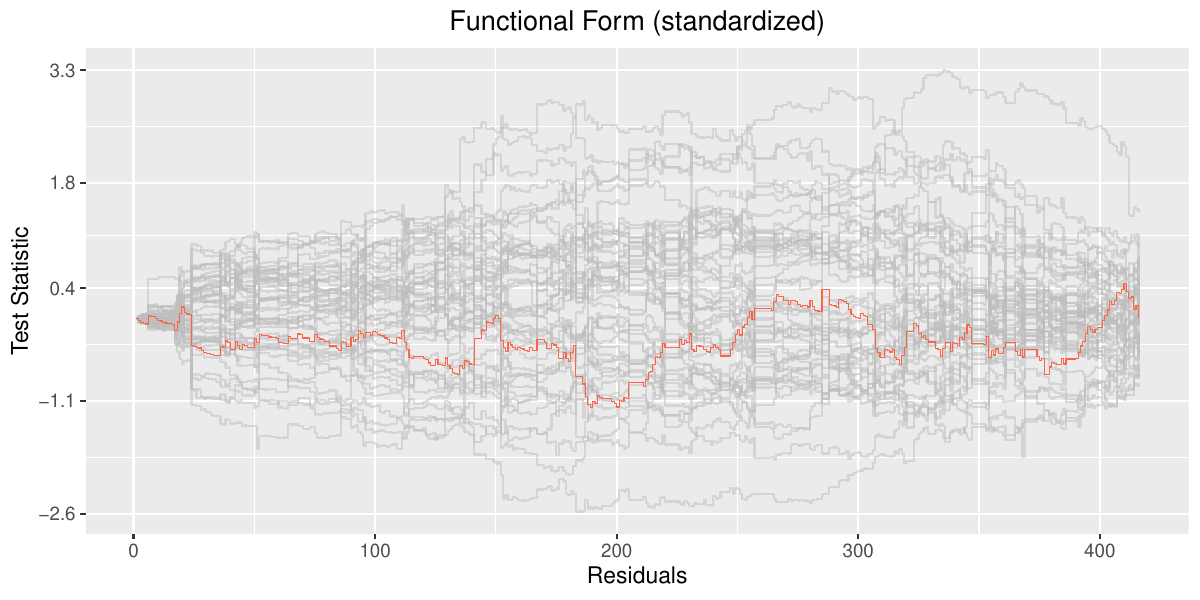}
    \caption{functional form test (\texttt{age})} \label{fig:pbc01_form4_ns_std}
  \end{subfigure}
  \caption{The sample path for the test statistic (red) overlays 50
    approximated sample paths under the null (grey) for
    Model~\eqref{eq:illustration:1}. Panel (a) is the plot for
    the omnibus test under the different quantiles of covariates. Panel (b) is 
    the link function test and panels (c)-(f) are functional form tests for 
    \texttt{bili}, \texttt{protime}, \texttt{albumin} and \texttt{age}, respectively.}
    \label{fig:pbc01:unstd}
\end{figure}

The output summarizes the key results, including the original function 
\code{Call} and the test's $p$-values. The unstandardized test yielded a 
non-significant $p$-value of 0.150, whereas the standardized test was 
statistically significant ($p$-value = 0.005). A significant result from the 
standardized test suggests potential violations of model assumptions, warranting 
further diagnostic analyses, such as link function and functional form tests. 
Calling the \code{plot()} method with \code{std = TRUE} generates 
Figure~\ref{fig:pbc01_omni_ns_std}, which is discussed alongside other 
diagnostic plots for Model~\ref{eq:illustration:1}.

To conduct the link function test for Model~\ref{eq:illustration:1}, we set
\code{testType = "link"}. This analysis utilized the same
\code{aftsrr()} object from the omnibus test along with the
non-smoothed Gehan-based estimating equations (\code{eqType =
  \code{"ns"}}).

\begin{knitrout}
\definecolor{shadecolor}{rgb}{0.969, 0.969, 0.969}\color{fgcolor}\begin{kframe}
\begin{alltt}
\hlkwd{system.time}\hldef{(}
  \hldef{pbc1_link_ns} \hlkwb{<-}
    \hlkwd{afttest}\hldef{(pbc1_aftsrr,} \hlkwc{data} \hldef{= pbc,} \hlkwc{npath} \hldef{=} \hlnum{200}\hldef{,} \hlkwc{testType} \hldef{=} \hlsng{"link"}\hldef{,}
            \hlkwc{estMethod} \hldef{=} \hlsng{"rr"}\hldef{,} \hlkwc{eqType} \hldef{=} \hlsng{"ns"}\hldef{,} \hlkwc{linApprox} \hldef{=} \hlnum{TRUE}\hldef{,} \hlkwc{seed} \hldef{=} \hlnum{0}\hldef{))}
\end{alltt}
\begin{verbatim}
##    user  system elapsed 
##   0.563   0.002   0.565
\end{verbatim}
\begin{alltt}
\hldef{pbc1_link_ns} \hlcom{# equivalent to print(pbc1_link_ns).}
\end{alltt}
\begin{verbatim}
## 
## 	AFT Goodness-of-Fit Test
## 
## Call: 
## afttest.aftsrr(object = pbc1_aftsrr, data = pbc, npath = 200, 
##     testType = "link", eqType = "ns", linApprox = TRUE, seed = 0, 
##     estMethod = "rr")
## 
## --- Null Hypothesis (H0) ---
## The relationship between covariates and the log survival time is 
## correctly specified. 
## 
## --- p-values ---
##  unstandardized standardized
##            0.10         0.02
\end{verbatim}
\end{kframe}
\end{knitrout}

\begin{knitrout}
\definecolor{shadecolor}{rgb}{0.969, 0.969, 0.969}\color{fgcolor}\begin{kframe}
\begin{alltt}
\hlkwd{plot}\hldef{(pbc1_link_ns,} \hlkwc{std} \hldef{=} \hlnum{TRUE}\hldef{)}
\end{alltt}
\end{kframe}
\end{knitrout}

%
%

While the unstandardized test was not statistically significant 
($p$-value = 0.100), the standardized test ($p$-value = 0.020) yielded a 
significant result. This outcome suggests potential violations of the model 
assumptions, warranting further investigation through functional form tests for 
each continuous covariate.

To check the functional form of each covariate under the
Model~\ref{eq:illustration:1}, we first defined the model using a
\code{Surv} object from the \pkg{survival} package. We then set the
following arguments for the analysis: \code{testType = "covForm"} was
used to specify the test type, and the rank-based estimator from the
\code{aftsrr()} function was selected with \code{estMethod = "rr"}. We
chose the non-smoothed Gehan-based estimating equations with
\code{eqType = \code{"ns"}}. A key requirement for this test is to
specify which covariate to examine using the \code{covTested}
argument. We set \code{covTested = "bili"} to specifically test
\code{bili}; otherwise, the function defaults to testing the first
covariate in the model formula.

\begin{knitrout}
\definecolor{shadecolor}{rgb}{0.969, 0.969, 0.969}\color{fgcolor}\begin{kframe}
\begin{alltt}
\hlkwd{system.time}\hldef{(}
  \hldef{pbc1_form1_ns} \hlkwb{<-} \hlkwd{afttest}\hldef{(}
      \hlkwd{Surv}\hldef{(time, status)} \hlopt{~} \hldef{bili} \hlopt{+} \hldef{protime} \hlopt{+} \hldef{albumin} \hlopt{+} \hldef{age} \hlopt{+} \hldef{edema,}
        \hlkwc{data} \hldef{= pbc,} \hlkwc{npath} \hldef{=} \hlnum{200}\hldef{,} \hlkwc{testType} \hldef{=} \hlsng{"covForm"}\hldef{,}
        \hlkwc{estMethod} \hldef{=} \hlsng{"rr"}\hldef{,} \hlkwc{eqType} \hldef{=} \hlsng{"ns"}\hldef{,} \hlkwc{covTested} \hldef{=} \hlsng{"bili"}\hldef{,}
        \hlkwc{linApprox} \hldef{=} \hlnum{TRUE}\hldef{,} \hlkwc{seed} \hldef{=} \hlnum{0}\hldef{))}
\end{alltt}
\begin{verbatim}
##    user  system elapsed 
##   0.627   0.004   0.630
\end{verbatim}
\begin{alltt}
\hldef{pbc1_form1_ns} \hlcom{# equivalent to print(pbc1_form1_ns).}
\end{alltt}
\begin{verbatim}
## 
## 	AFT Goodness-of-Fit Test
## 
## Call: 
## afttest.formula(object = Surv(time, status) ~ bili + protime + 
##     albumin + age + edema, data = pbc, npath = 200, testType = "covForm", 
##     estMethod = "rr", eqType = "ns", covTested = "bili", linApprox = TRUE, 
##     seed = 0)
## 
## --- Null Hypothesis (H0) ---
## The functional form for covariate 'bili' is correctly specified.
## 
## --- p-values ---
##  unstandardized standardized
##          <0.001       <0.001
\end{verbatim}
\end{kframe}
\end{knitrout}

\begin{knitrout}
\definecolor{shadecolor}{rgb}{0.969, 0.969, 0.969}\color{fgcolor}\begin{kframe}
\begin{alltt}
\hlkwd{plot}\hldef{(pbc1_form1_ns,} \hlkwc{std} \hldef{=} \hlnum{TRUE}\hldef{)}
\end{alltt}
\end{kframe}
\end{knitrout}

%
%

Both the unstandardized ($p$-value $< 0.001$) and standardized 
($p$-value $< 0.001$) tests produced significant results, indicating a potential 
misspecification in the functional form of \code{bili}. Therefore, in line with 
previous work by \citet{dickson1997primary} and \citet{jin2003rank}, we apply a 
log transformation to \code{bili} to create Model~\ref{eq:illustration:2}.

The generated Figure~\ref{fig:pbc01:unstd} using the \code{plot()} method with 
the option \code{std = TURE} presents the observed path in a red line alongside 
\code{50} standardized perturbed paths in grey lines. The diagnostic plots in 
Figures~\ref{fig:pbc01_omni_ns_std}–\ref{fig:pbc01_form1_ns_std} reveal a clear 
distinction in model fit. The omnibus test (Figure~\ref{fig:pbc01_omni_ns_std}), 
link function test (Figure~\ref{fig:pbc01_link_ns_std}), and the functional form 
test for \code{bili} (Figure~\ref{fig:pbc01_form1_ns_std}) all indicate potential 
model violations as the realized path of the standardized test statistic clearly 
deviates from the approximated null paths. Conversely, the functional form tests 
for \code{protime}, \code{albumin}, and \code{age} 
(Figures~\ref{fig:pbc01_form2_ns_std}–\ref{fig:pbc01_form4_ns_std}) show the 
observed paths remaining well within the null boundaries, suggesting an adequate 
fit for those covariates. Functional form tests were not conducted for the 
binary covariate, \code{edema}. The omnibus test plot specifically includes 
percentiles of $z$ (10\%, 25\%, 50\%, 75\%, and 90\%) used. For the omnibus test, 
the process value at the 10\% truncation point is almost zero. This behavior is 
a direct consequence of the weight function, $\pi_{i} \left( \bz \right) = I 
\left( \bZ_{i} \leq \bz \right)$, which evaluates to zero at this lower threshold.

Alternatively, users can obtain the goodness-of-fit results based on the 
least-squares estimation method by either providing a pre-fitted \code{aftgee()} 
object or directly passing the model formula to \code{afttest()} using the 
\code{afttest.formula} method, which performs both model fitting and testing in 
a single step. The
estimation method can be specified as \code{estMethod = "ls"}, which
utilizes the \code{aftgee()} function from the \proglang{R} package
\pkg{aftgee}~\citep{chiou2014fitting}. Note that when this estimation
method is used, the \code{eqType} argument is not applicable.

\subsection{Model~M2} \label{sec:illustration2}

We next evaluated Model~\ref{eq:illustration:2}, which was designed to
improve the fit of Model~\ref{eq:illustration:1} by applying a log
transformation to the covariate \code{bili}. The analyses for
Model~\ref{eq:illustration:2} were the same as for
Model~\ref{eq:illustration:1}.

As in Model~\ref{eq:illustration:1}, we fit the model first and then plug the 
\code{aftsrr()} object into the \code{afttest()} directly. 

\begin{knitrout}
\definecolor{shadecolor}{rgb}{0.969, 0.969, 0.969}\color{fgcolor}\begin{kframe}
\begin{alltt}
\hldef{pbc2_aftsrr} \hlkwb{<-} \hldef{aftgee}\hlopt{::}\hlkwd{aftsrr}\hldef{(}
  \hlkwd{Surv}\hldef{(time, status)} \hlopt{~} \hldef{log_bili} \hlopt{+} \hldef{protime} \hlopt{+} \hldef{albumin} \hlopt{+} \hldef{age} \hlopt{+} \hldef{edema,}
  \hlkwc{data} \hldef{= pbc,} \hlkwc{eqType} \hldef{=} \hlsng{"ns"}\hldef{,} \hlkwc{rankWeights} \hldef{=} \hlsng{"gehan"}\hldef{)}
\end{alltt}
\end{kframe}
\end{knitrout}

\begin{knitrout}
\definecolor{shadecolor}{rgb}{0.969, 0.969, 0.969}\color{fgcolor}\begin{kframe}
\begin{alltt}
\hlkwd{system.time}\hldef{(}
  \hldef{pbc2_omni_ns} \hlkwb{<-}
    \hlkwd{afttest}\hldef{(pbc2_aftsrr,} \hlkwc{data} \hldef{= pbc,} \hlkwc{npath} \hldef{=} \hlnum{200}\hldef{,} \hlkwc{testType} \hldef{=} \hlsng{"omnibus"}\hldef{,}
            \hlkwc{estMethod} \hldef{=} \hlsng{"rr"}\hldef{,} \hlkwc{linApprox} \hldef{=} \hlnum{TRUE}\hldef{,} \hlkwc{seed} \hldef{=} \hlnum{0}\hldef{))}
\end{alltt}
\begin{verbatim}
##    user  system elapsed 
##   1.226   0.142   1.375
\end{verbatim}
\begin{alltt}
\hldef{pbc2_omni_ns} \hlcom{# equivalent to print(pbc2_omni_ns).}
\end{alltt}
\begin{verbatim}
## 
## 	AFT Goodness-of-Fit Test
## 
## Call: 
## afttest.aftsrr(object = pbc2_aftsrr, data = pbc, npath = 200, 
##     testType = "omnibus", linApprox = TRUE, seed = 0, estMethod = "rr")
## 
## --- Null Hypothesis (H0) ---
## The assumed semiparametric AFT model fits the data adequately. 
## 
## --- p-values ---
##  unstandardized standardized
##           0.195        0.265
\end{verbatim}
\end{kframe}
\end{knitrout}

\begin{knitrout}
\definecolor{shadecolor}{rgb}{0.969, 0.969, 0.969}\color{fgcolor}\begin{kframe}
\begin{alltt}
\hlkwd{plot}\hldef{(pbc2_omni_ns)}
\end{alltt}
\end{kframe}
\end{knitrout}

%
%

The omnibus test for Model~\ref{eq:illustration:2} yields an unstandardized
$p$-value of 0.195 and a standardized $p$-value of 0.265. Since both
$p$-values are greater than the conventional 0.05 threshold for
significance, the test indicates that the model provides a good fit
for the data.

To perform the link function test for Model~\ref{eq:illustration:2}, we set 
\code{testType = "link"} and all other options remained consistent with those 
used in the omnibus test.

\begin{knitrout}
\definecolor{shadecolor}{rgb}{0.969, 0.969, 0.969}\color{fgcolor}\begin{kframe}
\begin{alltt}
\hlkwd{system.time}\hldef{(}
  \hldef{pbc2_link_ns} \hlkwb{<-}
    \hlkwd{afttest}\hldef{(pbc2_aftsrr,} \hlkwc{data} \hldef{= pbc,} \hlkwc{npath} \hldef{=} \hlnum{200}\hldef{,} \hlkwc{testType} \hldef{=} \hlsng{"link"}\hldef{,}
            \hlkwc{estMethod} \hldef{=} \hlsng{"rr"}\hldef{,} \hlkwc{linApprox} \hldef{=} \hlnum{TRUE}\hldef{,} \hlkwc{seed} \hldef{=} \hlnum{0}\hldef{))}
\end{alltt}
\begin{verbatim}
##    user  system elapsed 
##   0.589   0.076   0.672
\end{verbatim}
\begin{alltt}
\hldef{pbc2_link_ns} \hlcom{# equivalent to print(pbc2_link_ns).}
\end{alltt}
\begin{verbatim}
## 
## 	AFT Goodness-of-Fit Test
## 
## Call: 
## afttest.aftsrr(object = pbc2_aftsrr, data = pbc, npath = 200, 
##     testType = "link", linApprox = TRUE, seed = 0, estMethod = "rr")
## 
## --- Null Hypothesis (H0) ---
## The relationship between covariates and the log survival time is 
## correctly specified. 
## 
## --- p-values ---
##  unstandardized standardized
##           0.095        0.170
\end{verbatim}
\end{kframe}
\end{knitrout}

\begin{knitrout}
\definecolor{shadecolor}{rgb}{0.969, 0.969, 0.969}\color{fgcolor}\begin{kframe}
\begin{alltt}
\hlkwd{plot}\hldef{(pbc2_link_ns)}
\end{alltt}
\end{kframe}
\end{knitrout}

%
%

The link function test for Model~\ref{eq:illustration:2} produces an
unstandardized $p$-value of 0.095 and a standardized $p$-value of
0.170. As both values exceed the conventional 0.05 significance
threshold, the results suggest that the model fits the data well.

To test the functional form of the covariate, \code{"log_bili"}, we set 
\code{testType = "covForm"} and \code{covTested = "log_bili"}. We then defined 
the model with a \code{Surv} object from the \pkg{survival} package.

\begin{knitrout}
\definecolor{shadecolor}{rgb}{0.969, 0.969, 0.969}\color{fgcolor}\begin{kframe}
\begin{alltt}
\hlkwd{system.time}\hldef{(}
  \hldef{pbc2_form1_ns} \hlkwb{<-} \hlkwd{afttest}\hldef{(}
    \hlkwd{Surv}\hldef{(time, status)} \hlopt{~} \hldef{log_bili} \hlopt{+} \hldef{protime} \hlopt{+} \hldef{albumin} \hlopt{+} \hldef{age} \hlopt{+} \hldef{edema,}
      \hlkwc{data} \hldef{= pbc,} \hlkwc{npath} \hldef{=} \hlnum{200}\hldef{,} \hlkwc{testType} \hldef{=} \hlsng{"covForm"}\hldef{,}
      \hlkwc{estMethod} \hldef{=} \hlsng{"rr"}\hldef{,} \hlkwc{eqType} \hldef{=} \hlsng{"ns"}\hldef{,} \hlkwc{covTested} \hldef{=} \hlsng{"log_bili"}\hldef{,}
      \hlkwc{linApprox} \hldef{=} \hlnum{TRUE}\hldef{,} \hlkwc{seed} \hldef{=} \hlnum{0}\hldef{))}
\end{alltt}
\begin{verbatim}
##    user  system elapsed 
##   0.649   0.073   0.727
\end{verbatim}
\begin{alltt}
\hldef{pbc2_form1_ns} \hlcom{# equivalent to print(pbc2_form1_ns).}
\end{alltt}
\begin{verbatim}
## 
## 	AFT Goodness-of-Fit Test
## 
## Call: 
## afttest.formula(object = Surv(time, status) ~ log_bili + protime + 
##     albumin + age + edema, data = pbc, npath = 200, testType = "covForm", 
##     estMethod = "rr", eqType = "ns", covTested = "log_bili", 
##     linApprox = TRUE, seed = 0)
## 
## --- Null Hypothesis (H0) ---
## The functional form for covariate 'log_bili' is correctly specified.
## 
## --- p-values ---
##  unstandardized standardized
##           0.390        0.405
\end{verbatim}
\end{kframe}
\end{knitrout}

\begin{knitrout}
\definecolor{shadecolor}{rgb}{0.969, 0.969, 0.969}\color{fgcolor}\begin{kframe}
\begin{alltt}
\hlkwd{plot}\hldef{(pbc2_form1_ns)}
\end{alltt}
\end{kframe}
\end{knitrout}

%
%

The functional form test for \code{log_bili} produced similar results: the 
unstandardized version yielded a $p$-value of 0.390, while the standardized 
version gave 0.405; since neither is statistically significant, this supports 
the appropriateness of the log transformation for \code{bili}. 

\begin{figure}[tbp]
  \centering
  \begin{subfigure}{0.49\textwidth}
    \centering
    \captionsetup{justification=raggedright,singlelinecheck = true}
    \includegraphics[width=0.99\textwidth]{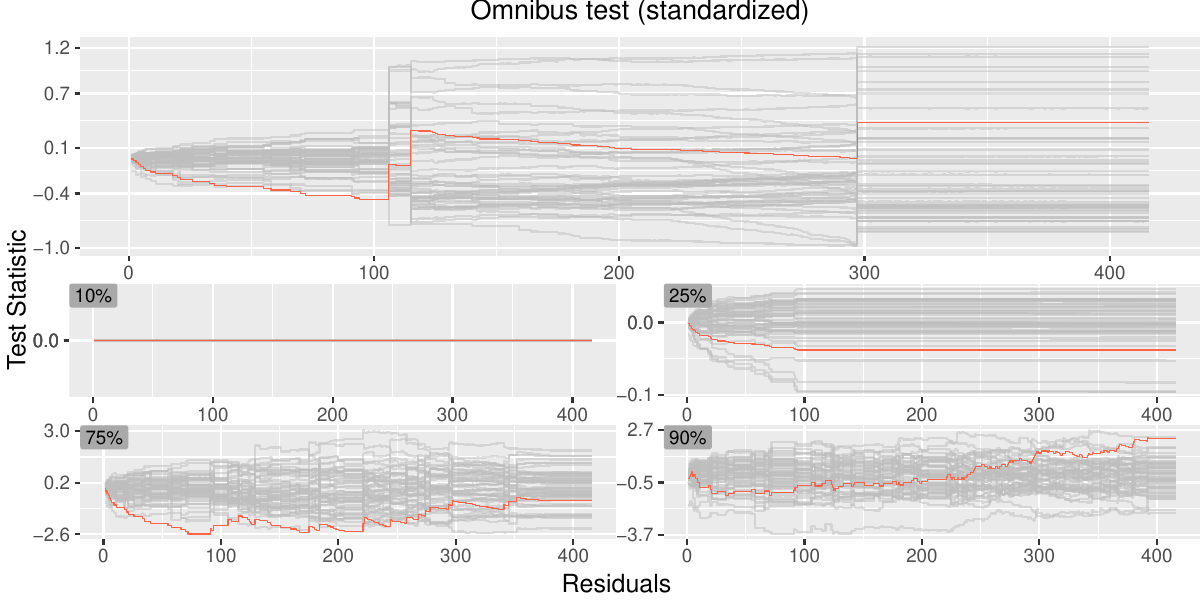}
    \caption{omnibus test } \label{fig:pbc02_omni_ns_std}
  \end{subfigure}
  \begin{subfigure}{0.49\textwidth}
    \centering
    \captionsetup{justification=raggedright,singlelinecheck = true}
    \includegraphics[width=0.99\textwidth]{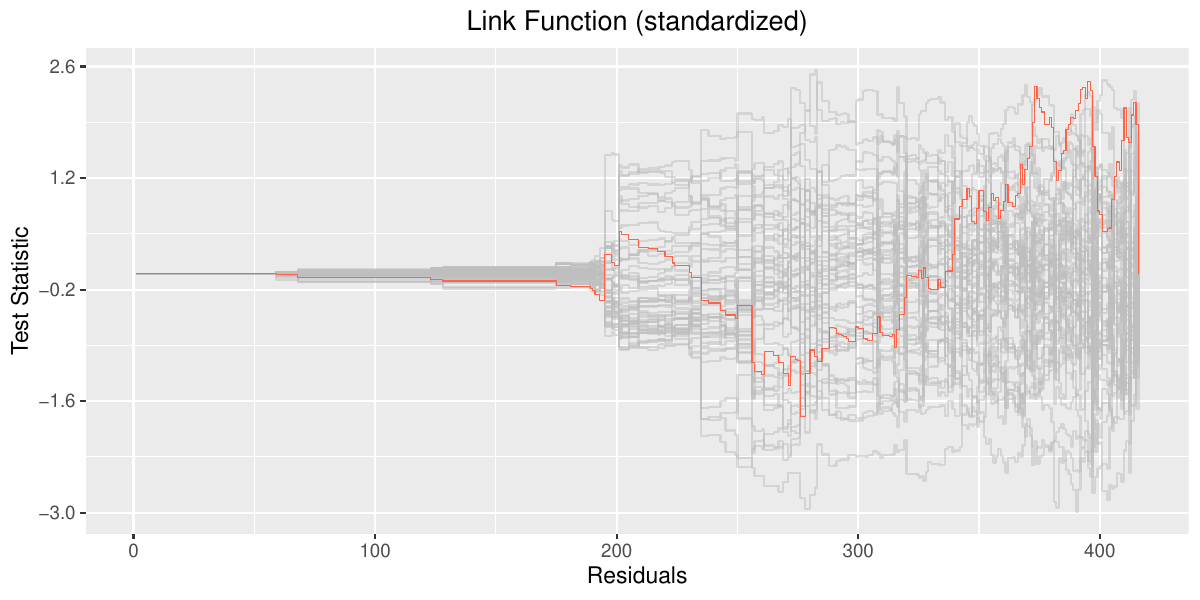}
    \caption{link function test} \label{fig:pbc02_link_ns_std}
  \end{subfigure}
  \\
  \begin{subfigure}{0.49\textwidth}
    \centering
    \captionsetup{justification=raggedright,singlelinecheck = true}
    \includegraphics[width=0.99\textwidth]{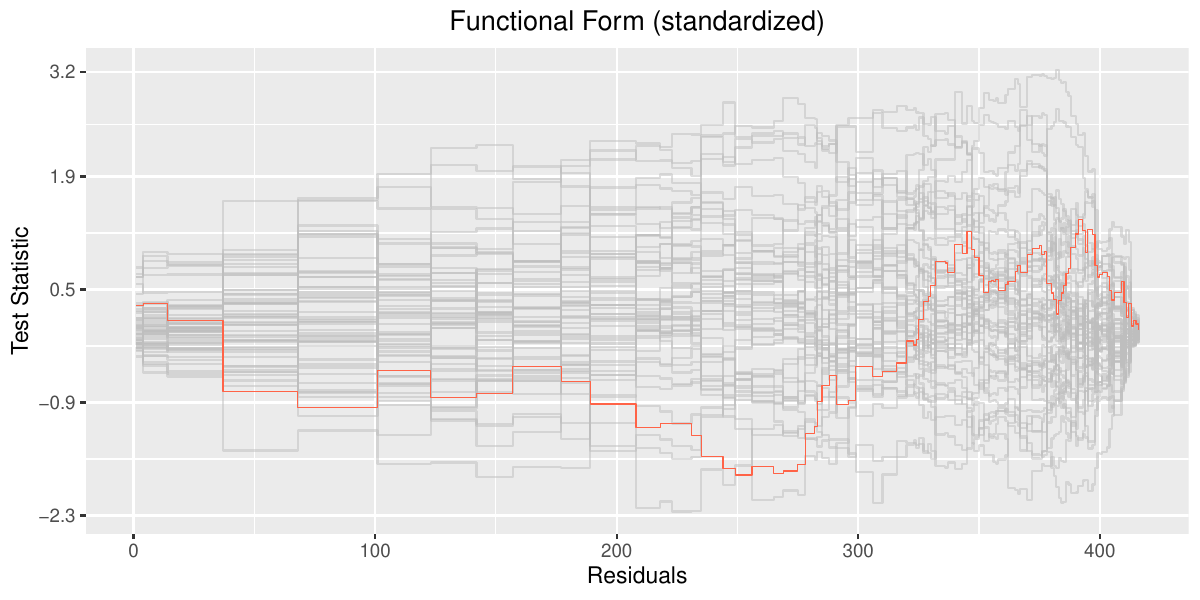}
    \caption{functional form test (\texttt{log\_bili})} \label{fig:pbc02_form1_ns_std}
  \end{subfigure}
  \begin{subfigure}{0.49\textwidth}
    \centering
    \captionsetup{justification=raggedright,singlelinecheck = true}
    \includegraphics[width=0.99\textwidth]{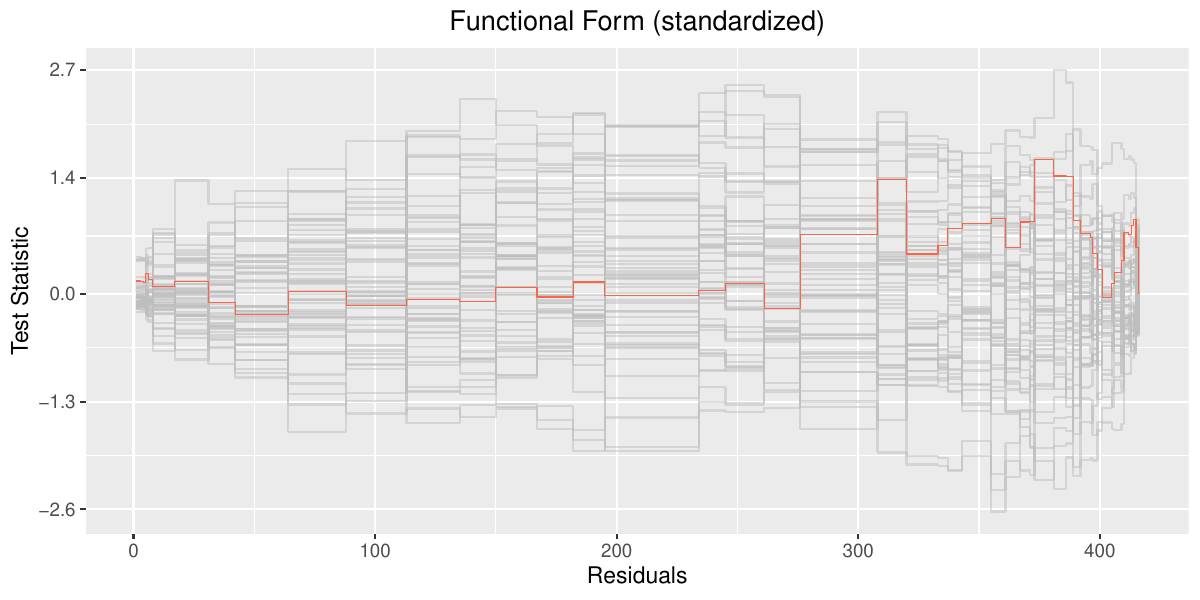}
    \caption{functional form test (\texttt{protime})} \label{fig:pbc02_form2_ns_std}
  \end{subfigure}
  \\
  \begin{subfigure}{0.49\textwidth}
    \centering
    \captionsetup{justification=raggedright,singlelinecheck = true}
    \includegraphics[width=0.99\textwidth]{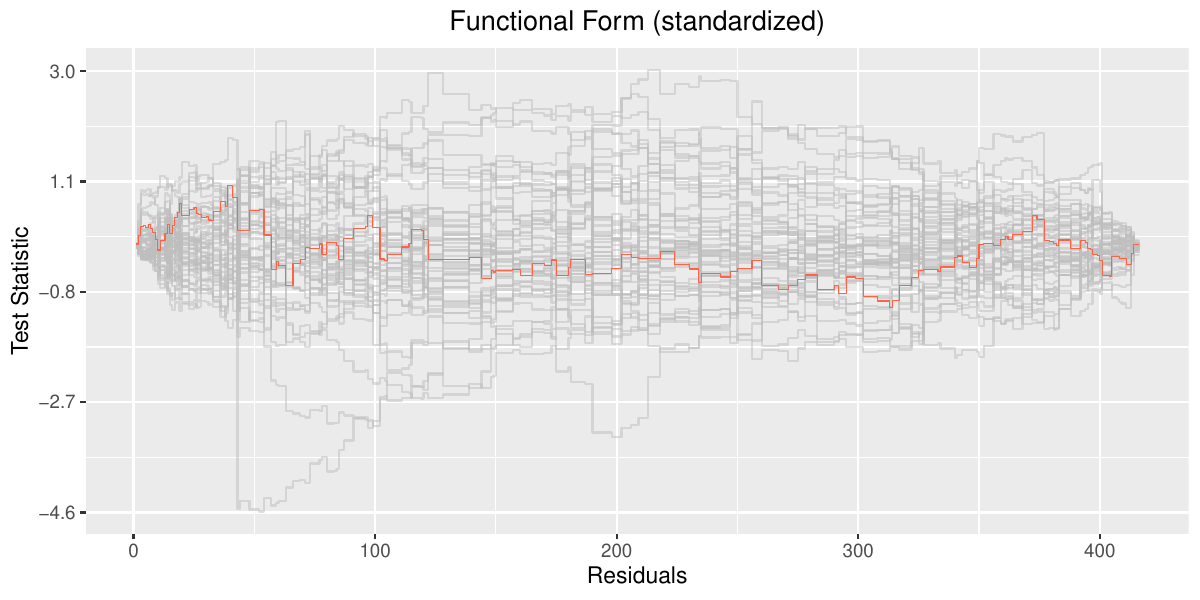}
    \caption{functional form test (\texttt{albumin})} \label{fig:pbc02_form3_ns_std}
  \end{subfigure}
  \begin{subfigure}{0.49\textwidth}
    \centering
    \captionsetup{justification=raggedright,singlelinecheck = true}
    \includegraphics[width=0.99\textwidth]{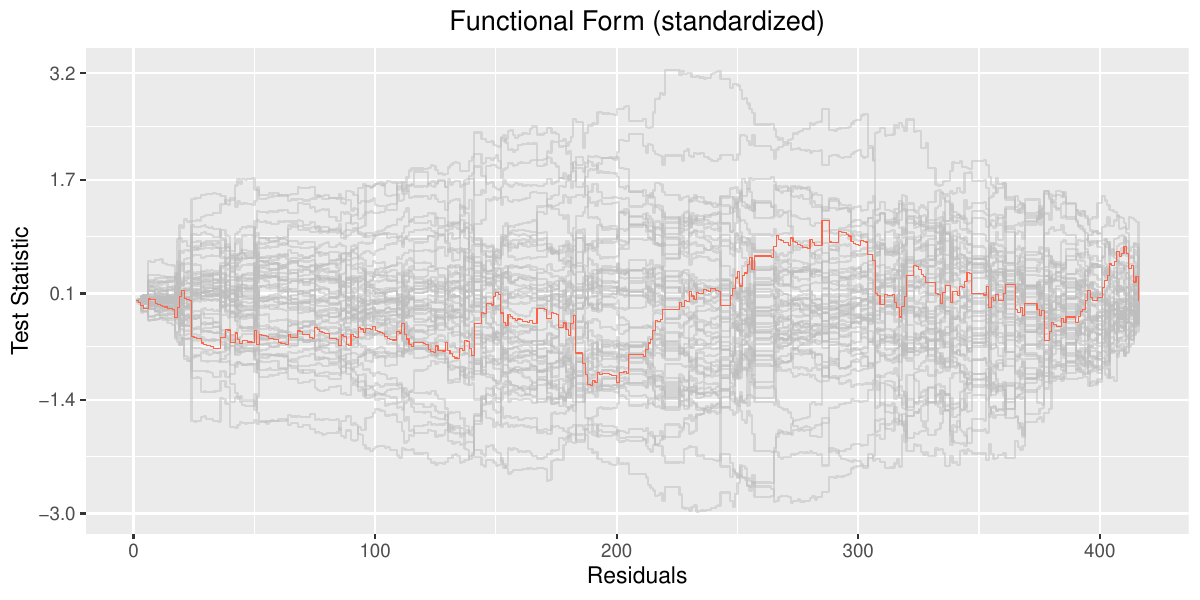}
    \caption{functional form test (\texttt{age})} \label{fig:pbc02_form4_ns_std}
  \end{subfigure}
  \caption{The sample path for the test statistic (red) overlays 50
    approximated sample paths under the null (grey) for
    Model~\eqref{eq:illustration:2} Panel (a) is the plot for the omnibus test 
    under the different quantiles of covariates. Panel (b) is the link function 
    test and panels (c)-(f) are functional form tests for \texttt{log\_bili}, 
    \texttt{protime}, \texttt{albumin} and \texttt{age}, respectively.}
\label{fig:pbc02:std}
\end{figure}

The results for Model~\ref{eq:illustration:2}, visualized using the \code{plot()} 
method with the default setting (\code{std = TRUE}), are shown in 
Figures~\ref{fig:pbc02_omni_ns_std}–\ref{fig:pbc02_form4_ns_std}. In all cases, 
the observed trajectories of the standardized test statistics lie largely within 
the 50 approximated sample paths generated under the null hypothesis, indicating 
no significant deviations from model assumptions and providing strong visual 
evidence that the semiparametric AFT model offers an adequate fit to the PBC data.

\section{Discussion} \label{sec:discussion}

The \pkg{afttest} package consolidates residual-based diagnostics 
for semiparametric AFT models into a unified and computationally 
efficient framework. It implements previously developed 
goodness-of-fit procedures while supporting both rank-based and 
least-squares estimators available in \pkg{aftgee}~\citep{chiou2014fitting}. 
For rank-based fits (\texttt{estMethod = "rr"}), users may select either 
the non-smoothed Gehan estimating equations 
(\texttt{eqType = "ns"}) or the induced-smoothing variant 
(\texttt{eqType = "is"}); for least-squares fits 
(\texttt{estMethod = "ls"}), the same diagnostic procedures apply. 
A central computational contribution is the implementation of an 
asymptotically equivalent linear-approximation strategy, which 
eliminates repeated numerical optimization in the multiplier 
bootstrap while preserving the limiting distribution of the test 
statistics. As a result, $p$-values based on Kolmogorov-type supremum 
statistics can be computed efficiently, and graphical displays of 
observed and perturbed stochastic paths are produced in a manner 
comparable to established Cox-model diagnostics~\citep{sas2016}.

The current implementation also establishes a scalable foundation 
for future extensions. Alternative estimators satisfying the 
regularity conditions of~\citep{choi2024general} could be incorporated, 
together with additional weight functions for the score process, 
including log-rank, Prentice--Wilcoxon, and the general $G^{\rho}$ 
class~\citep{chiou2015rank}, albeit with increased computational cost. 
Extensions to multivariate or time-varying AFT models and to settings 
with missing covariates would further expand applicability. The 
linear-approximation strategy developed here ensures that such 
extensions remain computationally tractable as model complexity 
and sample size increase.

\section*{Acknowledgments} \label{sec:acknowledgments}

This work was supported by the National Research Foundation of Korea (NRF) grant 
funded by the Korea government (MSIT [RS-2024-00341883]).

\bibliography{refs}

\clearpage
\appendix

\section{Multiplier Bootstrap Method} \label{sec:multiplier}
To approximate the null distribution of a complex test statistic, we
used a multiplier bootstrap approach. Unlike the standard bootstrap,
which resamples {\it data}, the multiplier bootstrap keeps the data
fixed and instead repeatedly perturbs the components of the test
statistic using independently generated random multipliers. The test statistic 
$\bW[t, \bz ; \bbh[n]][n]$ is defined as a weighted sum of martingale residuals. 
Under the regularity conditions C1--C7 established in Appendix by \citet{choi2024general}, 
its null distribution is approximated via a perturbed statistic, $\bWh[t, \bz; \bbh[n]][n]$. 
This perturbed version is constructed by weighting each subject's martingale 
residual with an independent random perturbation variable.

To detail the proposed multiplier bootstrap procedure, we first introduce a 
collection of essential notations and quantities. Let the counting and at-risk 
processes for the $i$-th subject be defined as $N_{i} \left( t, \bb \right) = 
I \left\{ e_{i} \left( \bb \right) \leq \log {t}, \Delta_{i} = 1 \right\}$ and 
$Y_{i} \left( t, \bb \right) = I \left\{ e_{i} \left( \bb \right) \geq \log {t} 
\right\}$, respectively. Building upon these individual processes, we introduce 
the aggregated empirical notation. For $d=0, 1$, let $S_{n}^{\left( d \right)} 
\left( t, \bb \right) = \sum_{i=1}^{n} \bZ_{i}^{\otimes d} Y_{i} \left( t, \bb 
\right)$ and define the ratio $E_{n} \left( t, \bb \right) = S_{n}^{\left( 1 
\right)} \left( t, \bb \right) / S_{n}^{\left( 0 \right)} \left( t, \bb \right)$. 
The cumulative hazard function $\Lambda \left( s, \bb \right)$ is estimated via 
the Nelson-Aalen estimator, $\widehat{\Lambda}_{n} \left( t; \bb \right) = 
\int_{0}^{t} J_{n} \left( s \right) \left\{ S_{n}^{\left( 0 \right)} \left( s; 
\bb \right) \right\}^{-1} d N_{n}^{\bullet} \left( s; \bb \right)$, where the 
aggregated counting process is given by $N_{n}^{\bullet} \left( s, \bb \right) 
= \sum_{i=1}^{n} N_{i} \left( t, \bb \right)$ with the risk set indicator 
$J_{n} \left( t \right) = I \left( S_{n}^{\left( 0 \right)} \left(t, \bb \right) 
> 0 \right)$. Consequently, the martingale residual for each subject is expressed 
as $\widehat{M}_{i} \left( t, \widehat{\bb}_{n} \right) = N_{i} \left( t, 
\widehat{\bb}_{n} \right) - \int_{0}^{t} Y_{i} \left( s, \widehat{\bb}_{n} 
\right) d \widehat{\Lambda}_{n} \left( s, \widehat{\bb}_{n} \right)$.

Regarding the weighted processes associated with $\pi(\cdot)$, we define 
$S_{n}^{\left( \pi \right)} \left(t, \bz, \bb \right) = \sum_{i=1}^{n} \pi_{i} \left( 
\bz \right) Y_{i} \left( t, \bb \right)$ along with $E_{n}^{\left( \pi \right)} 
\left(t, \bz, \bb \right) = S_{n}^{\left( \pi \right)} \left(t, \bz, \bb \right) 
/ S_{n}^{\left( 0 \right)} \left( t, \bb \right)$. In addition, let the weighted 
density processes be $f_{n}^{\left( \pi \right)} \left( t,\bz \right) = n^{-1} 
\sum_{i=1}^{n} \Delta_{i} \pi_{i} \left( \bz \right) f^{\left( 0 \right)} 
\left( t \right) t \bZ_{i}$ and $g_{n}^{\left( \pi \right)} \left( t,\bz \right) 
= n^{-1} \sum_{i=1}^{n} \pi_{i} \left( \bz \right) g^{\left( 0 \right)} 
\left( t \right) t \bZ_{i}$. Here, $f^{\left( 0 \right)} \left( t \right)$ and 
$g^{\left( 0 \right)} \left( t \right)$ correspond to the baseline densities of 
the transformed variables $T_{i} \exp \left( {\bZ_{i}^{\top} \bb_{0}} \right)$ 
and $X_{i} \exp \left( {\bZ_{i}^{\top} \bb_{0}} \right)$, respectively. The 
estimators $\widehat{f}_{n}^{\left( 0 \right)} \left( t \right)$ and 
$\widehat{g}_{n}^{\left( 0 \right)} \left( t \right)$ are derived utilizing 
kernel smoothing methods \citep{diehl1988kernel, novak2013goodness, 
silverman2018density}. Substitution of these kernel estimates yields 
$\widehat{f}_{n}^{\left( \pi \right)} \left( t, \bz \right)$ and 
$\widehat{g}_{n}^{\left( \pi \right)} \left( t, \bz \right)$; we refer to 
\citet{choi2024general} for their explicit expressions. 

Let $\left( \phi_{1}, \cdots, \phi_{n} \right)$ denote $n$ independent and 
identically distributed (i.i.d.) positive multiplier random variables such that 
$\mE[\phi_{i}] = \mvar{\phi_{i}} = 1$. It follows that each $\phi_{i} - 1$ is 
mean-zero with unit variance. Let 
\begin{gather*}
    \score[t, \bb][n] 
    = \sum_{i=1}^{n} \int_{0}^{t} \psi_{n} \left( s, \bb \right) \left\{ \bZ_{i} - E_{n} \left( s, \bb \right) \right\} \dd N_{i} \left(s, \bb \right) \\
    \score[t, \bb][n][\phi]
    = \sum_{i=1}^{n} \int_{0}^{t} \psi_{n} \left( s, \bb \right) \left\{ \bZ_{i} - E_{n} \left( s, \bb \right) \right\} d \Mh[s, \bb][n,i] \left( \phi_{i} - 1 \right), \mbox{ and } \\
    \score[t, \bz, \bb][n][\left( \pi \right) \phi] 
    = \sum_{i=1}^{n} \int_{0}^{t} \left\{ \pi_{i} \left( \bz \right) - E_{n}^{\left( \pi \right)} \left( s, \bz, \bb \right) \right\} d \Mh[s, \bb][n,i] \left( \phi_{i} - 1 \right). 
\end{gather*}

Then, \citet{choi2024general} defined $\bWh[t, \bz; \bbh[n]][n]$ as 
\begin{align*}
    \bWh[t, \bz; \bbh[n]][n] 
    &= 
    n^{-\frac{1}{2}} \score[t, \bz; \bbh[n] ][n ][\left( \pi \right) \phi] 
    - n^{\frac{1}{2}} \left\{ \widehat{f}_{n}^{\left( \pi \right)} \left( t, \bz \right) + \int_{0}^{t} \widehat{g}_{n}^{\left( \pi \right)} \left( s, \bz\right) d \widehat{\Lambda}_{n} \left(s; \bbh[n] \right) \right\}^{\top} \left( \bbh[n] - \bbh[n][\phi] \right) \\
    & \hspace{4em} 
    - n^{-\frac{1}{2}} \int_{0}^{t} S_{n}^{\left( \pi \right)} \left( s, \bz; \bbh[n] \right) d \left\{ \widehat{\Lambda}_{n} \left(s; \bbh[n] \right) - \widehat{\Lambda}_{n} \left(s; \bbh[n][\phi] \right) \right\} 
\end{align*}
where $\bbh[n][\phi]$ is the solution to $\score[\infty, \bb][n] = 
\score[\infty, \bbh[n]][n][\phi]$. With these definitions established, we now 
demonstrate that the computationally efficient statistic $\bWh[t, \bz; 
\bbh[n]][n][\dag]$ is asymptotically equivalent to the original theoretical 
statistic $\bWh[t, \bz; \bbh[n]][n]$.

The calculation of $\bWh[t, \bz; \bbh[n]][n]$ involves obtaining the perturbed 
estimator $\bbh[n][\phi]$ for each bootstrap iteration by solving $\score[\infty, 
\bb][n] = \score[\infty, \bbh[n]][n][\phi]$. This requirement creates a 
significant computational burden, particularly for large sample sizes or extensive 
resampling. To address this efficiency bottleneck, we propose a statistically 
equivalent test statistic, $\bWh[t, \bz; \bbh[n]][n][\dag]$, which relies on a 
linear approximation and thereby eliminates the need for iterative optimization 
during the resampling process.

We first note that the following asymptotic expansions hold 
\citep{lin1998accelerated, choi2024general}:
\begin{gather}
    n^{\frac{1}{2}} \left\{ \widehat{\Lambda}_{n} \left( t , \bb \right) - \widehat{\Lambda}_{n} \left( t , \bb_{0} \right) \right\} 
    = n^{\frac{1}{2}} \boldsymbol{\kappa_{n}}^{\top} \left( t \right) \left( \bb - \bb_{0} \right) + o_{p} \left( 1 \right), \tag{Result 1} \label{eq:result1} \\
    n^{\frac{1}{2}} \left( \bbh[n] - \bbh[n][\phi] \right) 
    = \boldsymbol{\Omega_{n}}^{-1} \score[\bbh[n][\phi]][n] + o_{p} \left( 1 \right) \tag{Result 2} \label{eq:result2}
\end{gather}
where
\begin{align*}
    \boldsymbol{\kappa_{n}} \left( t \right) = - \int_{0}^{t} E_{n} \left( s \right) d \left( \lambda_{0} \left( s \right) s \right) \mbox{ and }
    \boldsymbol{\Omega_{n}} = \frac{1}{n} \frac{\partial \score[\bb][n]}{\partial \bb}.
\end{align*}

Leveraging these expansions, we define the approximated test statistic $\bWh[t, \bz; \bbh[n]][n][\dag]$ as
\begin{align*}
    \bWh[t, \bz; \bbh[n]][n][\dag] 
    = 
    n^{-\frac{1}{2}} \score[t, \bz; \bbh[n] ][n ][\left( \pi \right) \phi] - 
    \bDh[t, \bz; \bbh[n]][n][\top] \boldsymbol{\widehat{\Omega}_{n}}^{-1} \score[\bbh[n]][n][\phi]
\end{align*}
where $\bDh[t, \bz; \bbh[n]][n]$ is defined as
\begin{gather*}
    \bDh[t, \bz; \bbh[n]][n] 
    = 
    \widehat{f}_{n}^{\left( \pi \right)} \left( t, \bz \right) + 
    \int_{0}^{t} \widehat{g}_{n}^{\left( \pi \right)} \left( s, \bz \right) d \widehat{\Lambda}_{n} \left(s; \bbh[n] \right) + 
    \int_{0}^{t} \frac{1}{n} S_{n}^{\left( \pi \right)} \left( s, \bz; \bbh[n] \right) d \widehat{\boldsymbol{\kappa_{n}}} \left( s \right).
\end{gather*}

Note that the estimated $i$th influence function in \eqref{sec:test:3}, denoted by $\hat{h}_{i} \left( t, \bz; \bbh[n] \right)$, can be expressed as
\begin{align*} 
    \hat{h}_{i} \left( t, \bz; \bbh[n] \right)
    &= \int_{0}^{t} \left\{ \pi_{i} \left( \bz \right) - E_{n}^{\left( \pi \right)} \left( s, \bz, \bb \right) \right\} d \Mh[s, \bb][n,i] \\   
    & \qquad - 
    n^{\frac{1}{2}} \bDh[t, \bz; \bbh[n]][n][\top] \boldsymbol{\widehat{\Omega}_{n}}^{-1} 
    \left\{ \int_{0}^{t} \psi_{n} \left( s, \bb \right) \left\{ \bZ_{i} - E_{n} \left( s, \bb \right) \right\} d \Mh[s, \bb][n,i] \right\}. 
\end{align*}

By applying a Taylor expansion to the original statistic $\bWh[t, \bz ; \bbh[n]][n]$, we have
\begin{align*}
    & \bWh[t, \bz ; \bbh[n]][n] - n^{-\frac{1}{2}} \score[t, \bz; \bbh[n] ][n ][\left( \pi \right) \phi] \\
    &= - n^{\frac{1}{2}} \left\{ \widehat{f}_{n}^{\left( \pi \right)} \left( t, \bz \right) + \int_{0}^{t} \widehat{g}_{n}^{\left( \pi \right)} \left( s, \bz\right) d \widehat{\Lambda}_{n} \left(s; \bbh[n] \right) \right\}^{\top} \left( \bbh[n] - \bbh[n][\phi] \right) \\
    & \hspace{3em} 
    - n^{-\frac{1}{2}} \int_{0}^{t} S_{n}^{\left( \pi \right)} \left( s, \bz; \bbh[n] \right) d \left\{ \widehat{\Lambda}_{n} \left(s; \bbh[n] \right) - \widehat{\Lambda}_{n} \left(s; \bbh[n][\phi] \right) \right\} \\
    &= - n^{\frac{1}{2}} \left\{ \widehat{f}_{n}^{\left( \pi \right)} \left( t, \bz \right) + \int_{0}^{t} \widehat{g}_{n}^{\left( \pi \right)} \left( s, \bz\right) d \widehat{\Lambda}_{n} \left(s; \bbh[n] \right) \right\}^{\top} \left( \bbh[n] - \bbh[n][\phi] \right)\\
    & \hspace{3em} - n^{\frac{1}{2}} \int_{0}^{t} \frac{1}{n}
        S_{n}^{\left( \pi \right)} \left( s, \bz; \bbh[n] \right) d
        \boldsymbol{\kappa_{n}} \left( s \right) \left( \bbh[n] -
        \bbh[n][\phi] \right) + o_{p}(1) \mbox{ by \eqref{eq:result1}} \\
    &= - n^{\frac{1}{2}} \bDh[t, \bz; \bbh[n]][n][* \top] \left( \bbh[n] - \bbh[n][\phi] \right) + o_{p}(1) \\
    &= - \bDh[t, \bz; \bbh[n]][n][* \top] \boldsymbol{\Omega_{n}}^{-1} \score[\bbh[n][\phi]][n] + o_{p}(1)  \mbox{ by \eqref{eq:result2}},
\end{align*}
where
\begin{align*}
    \bDh[t, \bz; \bbh[n]][n][*] 
    = \widehat{f}_{n}^{\left( \pi \right)} \left( t, \bz \right) + \int_{0}^{t} \widehat{g}_{n}^{\left( \pi \right)} \left( s, \bz\right) d \widehat{\Lambda}_{n} \left(s; \bbh[n] \right) + \int_{0}^{t} \frac{1}{n} S_{n}^{\left( \pi \right)} \left( s, \bz; \bbh[n] \right) d \boldsymbol{\kappa_{n}} \left( s \right).
\end{align*}

We observe that $\bWh[t, \bz ; \bbh[n]][n]$ shares the same asymptotic 
representation as $\bWh[t, \bz; \bbh[n]][n][\dag]$. By substituting the 
consistent estimators $\boldsymbol{\widehat{\kappa}}$ and 
$\boldsymbol{\widehat{\Omega}_{n}}$ for $\boldsymbol{\kappa_{n}}$ and 
$\boldsymbol{\Omega_{n}}$, respectively, we replace 
$\bDh[t, \bz; \bbh[n]][n][* \top]$ with $\bDh[t, \bz; \bbh[n]][n][\top]$ without 
affecting the asymptotic behavior. Furthermore, $\score[\bbh[n]][n][\phi]$ has 
the same limiting distribution as $\score[\bbh[n][\phi]][n]$. Consequently, 
$\bWh[t, \bz; \bbh[n]][n][\dag]$ exhibits identical limiting finite-dimensional 
distributions to $\bW[t, \bz ; \bbh[n]][n]$, and its tightness follows from the 
same arguments established for $\bW[t, \bz ; \bbh[n]][n]$.

\section{Detailed Simulation Results} \label{sec:detailed}
For simulation settings, we adopted the simulation framework outlined in Scenario 
2 of~\citet{choi2024general}. The failure times were generated from an AFT model 
incorporating both linear and quadratic effects of the covariate $Z{i}$. The data 
generating process is defined as:
\begin{align}
    \log T_{i} = - \beta_{0} - \beta_{1} Z_{i} - \gamma Z_{i}^{2} + \epsilon_{i}, \label{model:sim1}
\end{align}
where $Z_{i}$ follows a normal distribution with mean $2$ and standard deviation 
$1$, and the error term $\epsilon_{i}$ is drawn from a standard normal 
distribution. We fixed the regression coefficients at $\beta_{0} = -4$ and 
$\beta_{1} = 1$, while varying the quadratic coefficient $\gamma$ from $0$ to 
$0.5$ in increments of $0.1$. This configuration allows us to assess the Type 
\Romannum{1} error rate under the correct model specification ($\gamma = 0$) and 
evaluate the power of the test under model misspecification ($\gamma \neq 0$).

We considered three sample sizes ($n = 100, 300$ and $500$) with a target censoring 
rate of approximately 20\%. Censoring times were independently generated from a 
log-normal distribution with unit standard deviation and a mean $\tau$ calibrated 
to achieve the desired censoring proportion. For inference, $500$ approximated 
sample paths were generated from the null model to compute $p$-values for each 
scenario. The null hypothesis was rejected when the simulated $p$-value fell 
below the nominal significance level of $\alpha = 0.05$. We report results for 
both the unstandardized and standardized versions of the test statistics.
{
\begin{table}[htp]
    \fontsize{10}{11}\selectfont
    \centering
    \caption{\label{tab:sim:result} Rejection rates with the nominal level of 
    0.05 under the Model \eqref{model:sim1}. The omnibus (omni), link function 
    (link), and functional form (form) tests are considered using the non-smooth 
    estimator (ns), induced smoothed estimator (is), and the least-squares 
    estimator (ls) under the sample sizes of $n = 100, 300$, and $500$ with a 
    censoring rate of 20\%. Each number in a cell is based on 1,000 replications 
    for the standardized statistic (bold) or unstandardized statistic.} 
    \begin{tabular}{cccccccccccc}
        \toprule
        \multicolumn{3}{c}{ } & \multicolumn{3}{c}{N = 100} & \multicolumn{3}{c}{N = 300} & \multicolumn{3}{c}{N = 500} \\
        \cmidrule(l{3pt}r{3pt}){4-6} \cmidrule(l{3pt}r{3pt}){7-9} \cmidrule(l{3pt}r{3pt}){10-12}
        $\gamma$ & Test & std & ns & is & ls & ns & is & ls & ns & is & ls\\
        \midrule
         &  & T & \textbf{0.003} & \textbf{0.003} & \textbf{0.002} & \textbf{0.010} & \textbf{0.010} & \textbf{0.009} & \textbf{0.008} & \textbf{0.008} & \textbf{0.007}\\
        
         & \multirow[t]{-2}{*}{omni} & F & 0.002 & 0.002 & 0.001 & 0.003 & 0.003 & 0.000 & 0.004 & 0.004 & 0.004\\
        
         &  & T & \textbf{0.005} & \textbf{0.005} & \textbf{0.008} & \textbf{0.021} & \textbf{0.021} & \textbf{0.020} & \textbf{0.021} & \textbf{0.021} & \textbf{0.021}\\
        
         & \multirow[t]{-2}{*}{link} & F & 0.005 & 0.005 & 0.006 & 0.006 & 0.006 & 0.004 & 0.009 & 0.009 & 0.009\\
        
         &  & T & \textbf{0.012} & \textbf{0.012} & \textbf{0.010} & \textbf{0.009} & \textbf{0.009} & \textbf{0.012} & \textbf{0.014} & \textbf{0.014} & \textbf{0.014}\\
        
        \multirow[t]{-6}{*}{0} & \multirow[t]{-2}{*}{form} & F & 0.006 & 0.006 & 0.004 & 0.005 & 0.005 & 0.003 & 0.008 & 0.008 & 0.006\\
        
         &  & T & \textbf{0.002} & \textbf{0.002} & \textbf{0.003} & \textbf{0.052} & \textbf{0.052} & \textbf{0.065} & \textbf{0.108} & \textbf{0.108} & \textbf{0.100}\\
        
         & \multirow[t]{-2}{*}{omni} & F & 0.001 & 0.001 & 0.000 & 0.008 & 0.008 & 0.005 & 0.012 & 0.012 & 0.011\\
        
         &  & T & \textbf{0.018} & \textbf{0.018} & \textbf{0.018} & \textbf{0.181} & \textbf{0.181} & \textbf{0.182} & \textbf{0.246} & \textbf{0.246} & \textbf{0.247}\\
        
         & \multirow[t]{-2}{*}{link} & F & 0.004 & 0.004 & 0.002 & 0.013 & 0.013 & 0.011 & 0.021 & 0.021 & 0.024\\
        
         &  & T & \textbf{0.030} & \textbf{0.030} & \textbf{0.033} & \textbf{0.162} & \textbf{0.162} & \textbf{0.158} & \textbf{0.215} & \textbf{0.215} & \textbf{0.221}\\
        
        \multirow[t]{-6}{*}{0.1} & \multirow[t]{-2}{*}{form} & F & 0.002 & 0.002 & 0.002 & 0.010 & 0.010 & 0.012 & 0.020 & 0.020 & 0.022\\
        
         &  & T & \textbf{0.008} & \textbf{0.008} & \textbf{0.015} & \textbf{0.243} & \textbf{0.243} & \textbf{0.276} & \textbf{0.515} & \textbf{0.515} & \textbf{0.539}\\
        
         & \multirow[t]{-2}{*}{omni} & F & 0.006 & 0.006 & 0.007 & 0.025 & 0.025 & 0.029 & 0.069 & 0.069 & 0.104\\
        
         &  & T & \textbf{0.054} & \textbf{0.054} & \textbf{0.064} & \textbf{0.480} & \textbf{0.480} & \textbf{0.479} & \textbf{0.697} & \textbf{0.697} & \textbf{0.714}\\
        
         & \multirow[t]{-2}{*}{link} & F & 0.010 & 0.010 & 0.016 & 0.064 & 0.064 & 0.075 & 0.118 & 0.118 & 0.149\\
        
         &  & T & \textbf{0.095} & \textbf{0.095} & \textbf{0.101} & \textbf{0.488} & \textbf{0.488} & \textbf{0.503} & \textbf{0.708} & \textbf{0.708} & \textbf{0.726}\\
        
        \multirow[t]{-6}{*}{0.2} & \multirow[t]{-2}{*}{form} & F & 0.008 & 0.008 & 0.012 & 0.057 & 0.057 & 0.054 & 0.119 & 0.119 & 0.154\\
        
         &  & T & \textbf{0.018} & \textbf{0.018} & \textbf{0.036} & \textbf{0.470} & \textbf{0.470} & \textbf{0.494} & \textbf{0.862} & \textbf{0.862} & \textbf{0.871}\\
        
         & \multirow[t]{-2}{*}{omni} & F & 0.002 & 0.002 & 0.007 & 0.054 & 0.054 & 0.082 & 0.189 & 0.189 & 0.248\\
        
         &  & T & \textbf{0.089} & \textbf{0.089} & \textbf{0.096} & \textbf{0.685} & \textbf{0.685} & \textbf{0.703} & \textbf{0.932} & \textbf{0.932} & \textbf{0.939}\\
        
         & \multirow[t]{-2}{*}{link} & F & 0.014 & 0.014 & 0.023 & 0.123 & 0.123 & 0.160 & 0.282 & 0.282 & 0.353\\
        
         &  & T & \textbf{0.131} & \textbf{0.131} & \textbf{0.158} & \textbf{0.755} & \textbf{0.755} & \textbf{0.764} & \textbf{0.957} & \textbf{0.957} & \textbf{0.957}\\
        
        \multirow[t]{-6}{*}{0.3} & \multirow[t]{-2}{*}{form} & F & 0.008 & 0.008 & 0.019 & 0.139 & 0.139 & 0.185 & 0.356 & 0.356 & 0.418\\
        
         &  & T & \textbf{0.029} & \textbf{0.029} & \textbf{0.048} & \textbf{0.694} & \textbf{0.694} & \textbf{0.734} & \textbf{0.975} & \textbf{0.975} & \textbf{0.972}\\
        
         & \multirow[t]{-2}{*}{omni} & F & 0.002 & 0.002 & 0.010 & 0.118 & 0.118 & 0.180 & 0.386 & 0.386 & 0.457\\
        
         &  & T & \textbf{0.111} & \textbf{0.111} & \textbf{0.125} & \textbf{0.824} & \textbf{0.824} & \textbf{0.837} & \textbf{0.992} & \textbf{0.992} & \textbf{0.989}\\
        
         & \multirow[t]{-2}{*}{link} & F & 0.018 & 0.018 & 0.039 & 0.215 & 0.215 & 0.289 & 0.534 & 0.534 & 0.605\\
        
         &  & T & \textbf{0.188} & \textbf{0.188} & \textbf{0.250} & \textbf{0.935} & \textbf{0.935} & \textbf{0.938} & \textbf{0.996} & \textbf{0.996} & \textbf{0.996}\\
        
        \multirow[t]{-6}{*}{0.4} & \multirow[t]{-2}{*}{form} & F & 0.015 & 0.015 & 0.036 & 0.286 & 0.286 & 0.342 & 0.652 & 0.652 & 0.733\\
        
         &  & T & \textbf{0.054} & \textbf{0.054} & \textbf{0.087} & \textbf{0.835} & \textbf{0.835} & \textbf{0.854} & \textbf{0.996} & \textbf{0.996} & \textbf{0.996}\\
        
         & \multirow[t]{-2}{*}{omni} & F & 0.002 & 0.002 & 0.024 & 0.194 & 0.194 & 0.272 & 0.617 & 0.617 & 0.687\\
        
         &  & T & \textbf{0.154} & \textbf{0.154} & \textbf{0.205} & \textbf{0.925} & \textbf{0.925} & \textbf{0.929} & \textbf{0.998} & \textbf{0.998} & \textbf{0.998}\\
        
         & \multirow[t]{-2}{*}{link} & F & 0.024 & 0.024 & 0.060 & 0.310 & 0.310 & 0.394 & 0.720 & 0.720 & 0.784\\
        
         &  & T & \textbf{0.315} & \textbf{0.315} & \textbf{0.393} & \textbf{0.981} & \textbf{0.981} & \textbf{0.978} & \textbf{1.000} & \textbf{1.000} & \textbf{1.000}\\
        
        \multirow[t]{-6}{*}{0.5} & \multirow[t]{-2}{*}{form} & F & 0.025 & 0.025 & 0.062 & 0.470 & 0.470 & 0.559 & 0.861 & 0.861 & 0.898\\
        \bottomrule
    \end{tabular}
\end{table}
}
Table~\ref{tab:sim:result} presents the results for the omnibus, link function, 
and functional form tests. 

We first evaluate the Type \Romannum{1} error rates under the null hypothesis 
($\gamma = 0$). While the empirical error rates are consistently below the 
nominal $0.05$ level, all tests appear conservative, with rates generally 
remaining at or below $0.025$. Similar behavior has been reported by 
\citet{novak2013goodness} and \citet{choi2024general}. Under misspecification 
($\gamma \neq 0$), power increases with both sample size and the magnitude of 
$\gamma$. Notably, standardized tests consistently outperform their 
unstandardized counterparts, demonstrating greater sensitivity to departures from 
the assumed AFT model. For instance, with $\gamma = 0.3$ and $n=300$, the 
least-squares functional form test yields a rejection rate of $0.764$ for the 
standardized version, compared to only $0.185$ for the unstandardized version. 
The proposed statistics perform comparably to rank-based methods (non-smooth 
and induced-smoothed). Furthermore, the functional form test exhibits higher 
power than the link function test, consistent with the findings of \citet{choi2024general}.

At small sample sizes ($n=100$), rejection rates are notably lower. For example, 
with $\gamma = 0.3$, our proposed standardized non-smooth functional form test 
yields a rejection rate of $0.158$ (unstandardized: $0.019$). In contrast, 
\citet{choi2024general} reports a higher rate of $0.342$ (unstandardized: $0.023$) 
under the same conditions. However, at larger sample sizes ($n=500$) with 
$\gamma = 0.3$, the performance gap narrows significantly. Our standardized test 
achieves a rejection rate of $0.957$ (unstandardized: $0.356$), which is 
comparable to the $0.971$ (unstandardized: $0.453$) reported by \citet{choi2024general}. 
Consequently, while \citet{choi2024general}'s method yields higher power in small 
samples, our proposed linear approximation becomes the preferred choice for larger 
sample sizes, such as $n \geq 500$, providing comparable power with substantially 
reduced computational cost. To facilitate practical application, we have 
implemented both methods in our software package. This allows researchers to 
leverage the statistical power of \citet{choi2024general}'s method for smaller 
datasets, while utilizing our efficient linear approximation for large-scale 
analyses where computational feasibility is paramount.

\end{document}